\documentclass[journal]{IEEEtran}

\RequirePackage[l2tabu, orthodox, abort]{nag}
\usepackage{cite}
\usepackage{array}
\usepackage{tabu}
\usepackage{booktabs}
\usepackage{comment}
\usepackage[cmex10]{amsmath}
\usepackage{amssymb}
\usepackage{bm}
\usepackage{color}
\usepackage{parskip}
\usepackage{textcomp}
\usepackage{siunitx}
\usepackage{graphicx}
\usepackage{algorithm}
\usepackage{algpseudocode}
\usepackage{lipsum}
\usepackage{fixltx2e}
\usepackage{contour}
\usepackage{pgf}
\pgfrealjobname{Tilley_TMI}
\DeclareMathOperator{\mD}{D}
\DeclareMathOperator{\mR}{R}
\DeclareMathOperator{\sinc}{sinc}

\DeclareMathOperator{\Poisson}{Poisson}
\DeclareMathOperator{\bias}{bias}
\DeclareMathOperator{\noise}{noise}

\DeclareMathOperator{\jaccard}{Jaccard}
\DeclareMathOperator{\mJac}{mJac}
\DeclareMathOperator*{\argmin}{argmin}
\newcommand{\fat}{0.01875}
\newcommand{\bone}{0.06044}
\newcommand{\avgfatbone}{0.039595}

\newcommand{\uCT}{\textmu{}CT}
\DeclareSIUnit\pixel{pixel}
\DeclareSIUnit\voxel{voxel}
\DeclareSIUnit\photon{photons}
\newcommand{\nlpwls}{GPL-BC}
\newcommand{\nlpwlsnc}{GPL-B}
\newcommand{\nlpwlsi}{GPL-I}
\newcommand{\fdkdeblur}{dFDK}
\newcommand{\SIr}[3]{\SI[round-mode=places, round-precision=#1]{#2}{#3}}
\newcommand{\numr}[2]{\num[round-mode=places, round-precision=#1]{#2}}

\newcommand{\vect}[1]{\bm{#1}}
\newcommand{\mat}[1]{\bm{\mathrm{#1}}}
\newcommand{\mx}{\mathrm{e}^{-\mat{A} \vect{\mu}}}

\usepackage{realdata}
\usepackage{todonotes}

\usepackage{ifthen}
\providecommand{\diffmode}{false}

\colorlet{diffcolor}{black}
\ifthenelse{\equal{\diffmode}{true}}{\colorlet{diffcolortwo}{blue}}{\colorlet{diffcolortwo}{black}}
\newenvironment{DIFmanualadd}{}{}
\newenvironment{DIFtwo}{\color{diffcolortwo}}{}

\providecommand{\buildappendices}{true}

\begin{document}

\title{Penalized-Likelihood Reconstruction with High-Fidelity Measurement Models for High-Resolution Cone-Beam Imaging}

\author{{Steven~Tilley~II,
Matthew~Jacobson, Qian~Cao, Michael~Brehler,
Alejandro~Sisniega, Wojciech~Zbijewski, J.~Webster~Stayman,~\IEEEmembership{Senior~Member,~IEEE}}%
\thanks{Copyright (c) 2017 IEEE. Personal use of this material is permitted. However, permission to use this material for any other purposes must be obtained from the IEEE by sending a request to pubs-permissions@ieee.org. The final version of this paper is S. Tilley et al., ``Penalized-Likelihood Reconstruction with High-Fidelity Measurement Models for High-Resolution Cone-Beam Imaging,'' IEEE Transactions on Medical Imaging, vol. PP, no. 99, pp. 1–1, 2017. DOI: 10.1109/TMI.2017.2779406 Available at http://ieeexplore.ieee.org/document/8125700/.
}\thanks{Department of Biomedical Engineering, Johns Hopkins University. email:~web.stayman@jhu.edu.}}

\maketitle

\begin{abstract} We present a novel reconstruction algorithm based on a general cone-beam CT
forward model which is capable of incorporating the blur and noise
correlations that are exhibited in flat-panel CBCT measurement data. Specifically, the proposed model 
may include scintillator blur, focal-spot blur,
and noise correlations due to light spread in the scintillator.
The proposed algorithm (\nlpwls{}) uses a Gaussian Penalized-Likelihood objective function
which incorporates models of Blur and Correlated noise.
In a simulation study, \nlpwls{} was able to achieve lower bias as compared to 
\begin{DIFmanualadd}deblurring followed by FDK as well as
a model-based reconstruction method without integration of measurement blur.\end{DIFmanualadd}
In the same study, \nlpwls{} was able to achieve \begin{DIFmanualadd}better line-pair reconstructions (in terms of segmented-image accuracy) as compared to deblurring followed by FDK, a model based method without blur,
and a model based method with blur but not noise correlations.\end{DIFmanualadd}
A prototype extremities quantitative cone-beam CT test bench
	was used to image a physical sample of human trabecular bone. These data were used to compare reconstructions using \begin{DIFmanualadd}the proposed method and model based methods without blur and/or correlation\end{DIFmanualadd}
to a registered \uCT\ image of the same bone sample.
The \nlpwls\ reconstructions resulted in more accurate
	trabecular bone segmentation.
	Multiple trabecular bone metrics, including Trabecular Thickness (Tb.Th.)  were computed for each reconstruction approach as well as the \uCT\ volume.
The \nlpwls\ reconstruction provided the most accurate
	Tb.Th.\ measurement, \SIr{3}{\nlpwlsspsmean}{\milli\meter}, as compared to the
\uCT\ derived value of \SIr{3}{\uCTmean}{\milli\meter}, followed by the \begin{DIFmanualadd}\nlpwlsnc{} 
	reconstruction, the \nlpwlsi{} reconstruction,  and then the FDK reconstruction (\SIr{3}{\nlpwlsspsBthreemean}{\milli\meter}, \SIr{3}{\nlpwlsspsImean}{\milli\meter},\end{DIFmanualadd} and
\SIr{3}{\fdkmean}{\milli\meter}, respectively).
\end{abstract}


\section{Introduction}

\IEEEPARstart{F}{lat}-panel-based cone-beam CT (CBCT) has offered more compact systems and improved spatial
resolution as compared to multirow detector CT (MDCT). These advantages have resulted in prototype and commercial CBCT systems
for specific applications, such as mammography~\cite{Lai2007, Kwan2007} and extremities imaging~\cite{Carrino2014, marinetto_quantification_2016},  %
where high spatial resolution is critical.
For example in mammography, clinicians would like to detect and visualize small
microcalcifications of \SI{<100}{\micro\meter}~\cite{Gong2004}. In extremities imaging, 
analysis of trabecular bone morphology for quantitative assessment is desired with trabecular detail of
\SIrange{50}{150}{\micro\meter}~\cite{Griffith2011}. The spatial resolution requirements for these tasks often
lie just beyond current system capabilities
(\SIrange{\sim180}{350}{\micro\meter}~\cite{baba_using_2004, bamba_image_2013} for
commercial systems).
Thus, even a modest improvement in spatial resolution has the potential to dramatically improve the clinical
utility of CBCT systems. 

Model-based iterative reconstruction (MBIR) techniques have been shown to
improve image quality in multi-detector CT (MDCT)~\cite{Thibault2007} as compared
to analytical approaches such as FDK~\cite{Feldkamp:84}. Much of the advantage
of MBIR methods derives from the inclusion of a high-fidelity forward model containing
both a model of the physical acquisition process and
a mathematical formulation of measurement statistics. For example, the noise model
informs the reconstruction algorithm about the relative information content of different
measurements, allowing weights on the relative importance of these measurements
in reconstructing the image.

While MBIR methods have been successfully applied to CBCT~\cite{Wang2014,
Dang2015, Sun2015}, the system models
are often borrowed directly from MDCT, and are therefore
derived from assumptions that may not be valid for CBCT. For example, MDCT detectors typically
include mechanisms to avoid signal sharing between detector elements (e.g., a pixelated scintillator)  
whereas flat-panel detectors typically exhibit significant sharing of the light generated by the primary x-ray
to secondary light quanta conversion. This effect can be prominent for smaller pixel sizes and leads to
increased blur and noise correlation between
neighboring measurements as compared to MDCT. 
While previous work\begin{DIFmanualadd}~\cite{Hofmann2013, hofmann_effects_2014}\end{DIFmanualadd} has suggested that focal spot modeling has relatively small advantages in current MDCT systems, the X-ray tubes used in
many dedicated CBCT systems tend to have stationary anodes and larger focal
spots than those used in MDCT.  %
Additionally, CBCT detectors have smaller pixels than MDCT.
Specifically,~\begin{DIFmanualadd}\cite{hofmann_effects_2014}\end{DIFmanualadd} demonstrated that focal spot modeling lead to improvements when the effective focal spot blur \begin{DIFmanualadd}(at the detector)\end{DIFmanualadd} was about \begin{DIFmanualadd}\num{1.3}\end{DIFmanualadd} detector elements wide,
which is an uncommon \begin{DIFmanualadd}occurrence\end{DIFmanualadd} in MDCT but common in CBCT (e.g.,~\begin{DIFmanualadd}\SI{0.2}{\milli\meter}\end{DIFmanualadd} pixels with a \SI{0.3}{\milli\meter} focal spot and a system magnification of 2).
Hence, focal spot blurring effects can be
significant in CBCT, particularly in systems that leverage higher magnifications. Traditional MDCT methods
do not incorporate such physical effects into their forward models, limiting their ability
to resolve fine resolution details when applied to flat-panel CBCT data. To get the most of such data (e.g., 
increasing spatial resolution capabilities), the 
MBIR forward model must adopt high-fidelity models of
these physical effects which are conventionally ignored.

Emission imaging has utilized high-fidelity modeling to recover lost spatial resolution for decades~\cite{Tsui1987, qi_fully_1998, qi_high-resolution_1998, formiconi_compensation_1989,
Chun2012, alessio_modeling_2006}.
The linear forward model in SPECT and PET imaging permits incorporation of
advanced blur models directly into the system matrix.
The resulting high-fidelity forward models are linear, simplifying optimization.
Such approaches have been used to model position-dependent geometric blurs and blurs due to physical detector characteristics (e.g., signal penetration through septa)~\cite{Tsui1987,
qi_fully_1998, qi_high-resolution_1998, formiconi_compensation_1989, Chun2012, alessio_modeling_2006}\begin{DIFmanualadd}.\end{DIFmanualadd}
Additionally, noise correlations induced by rebinning have been modeled for PET~\cite{Alessio2003}.

In contrast to emission imaging, transmission imaging forward models are fundamentally nonlinear due to the Beer-Lambert law,
preventing blurs from simply being incorporated into the system matrix.
Applications of advanced forward models in transmission tomography can be coarsely grouped into three
categories: sinogram restoration, direct MBIR, and preprocessing + MBIR. In
sinogram restoration approaches, ideal measurements or line integrals are
estimated based on a forward model. Images are reconstructed from these
estimates with either analytical methods (e.g., FDK) or MBIR with a simple forward model.
Sinogram restoration has been used in conjunction with models of blur~\cite{LaRiviere2006, LaRiviere2007} and noise correlation~\cite{Zhang2014}.  %
Direct MBIR methods incorporate the advanced forward models (e.g.,
of blur, noise correlation) directly into the MBIR objective function.
Such approaches have been used with an independent noise assumption~\cite{Yu2000, Feng2006}.  %
\begin{DIFmanualadd}Additionally, direct MBIR has been used to model blur and noise correlation in tomosynthesis~\cite{zheng_Detector_2017} by assuming
uniform quantum noise per view and that features of interest (e.g., microcalcifications) have low-attenuation and are small.\end{DIFmanualadd}  %
Preprocessing + MBIR is a hybrid approach, with the effects of preprocessing
modeled in the subsequent MBIR. For
example, noise correlations induced by deblurring have been
included in an MBIR model for CBCT~\cite{TilleyII2015a}.

Recently, we presented a novel preprocessing + MBIR method which can incorporate
spatial blur and noise correlations in a linear penalized weighted least-squares
framework~\cite{TilleyII2015a}.
This method demonstrates the importance of high fidelity system
modeling, specifically regarding spatial noise correlation, but contained
undesirable complexities due to the linearization of the forward model.
Specifically, this linearized forward model operates on an estimation of the
line integrals, requiring a preprocessing step that deconvolves system blurs
prior to reconstruction. The deconvolution requires solving an inverse problem
with tunable parameters such as regularization type and strength. To overcome
this limitation, we have presented a direct MBIR method with a non-linear
objective function and steepest-descent optimizer~\cite{tilley_nonlinear_2016},
which uses the measurement data directly to avoid the separate deblurring step and
resulting regularization of~\cite{TilleyII2015a}.

In this work we present a novel direct MBIR method based on the non-linear
least-squares forward model and objective function of~\cite{tilley_nonlinear_2016} which may incorporate
blur, noise correlation, and a Gaussian noise model for the measurements. The
optimization algorithm utilizes optimization transfer and separable surrogates, similar to the algorithm in~\cite{Erdogan1999, erdogan1999a}.
We derive the algorithm for this
Gaussian Penalized-Likelihood (PL) objective with modeled Blur and noise Correlation (\nlpwls), and evaluate performance
relative to the \begin{DIFmanualadd}same algorithm with simpler forward models\end{DIFmanualadd}.
Specifically, \nlpwlsi{} assumed there was no blur in the model (Identity blur), and \nlpwlsnc{} assumed no noise correlation.
\begin{DIFmanualadd}The GPL methods are compared to a Deblurring + FDK (\fdkdeblur{}) method, where measurement data are deblurred prior to FDK reconstruction.
Blur measurements from a prototype extremities quantitative CBCT (qCBCT) test bench~\cite{marinetto_quantification_2016} were used to construct a simulation study
measuring the image quality of reconstructed line-pairs.\end{DIFmanualadd}
The qCBCT test bench was used to
scan a sample of human trabecular bone. Reconstructions of the bone
using FDK, \begin{DIFmanualadd}\nlpwlsi{}, \nlpwlsnc{}\end{DIFmanualadd}, and \nlpwls{} were compared to each other as well as a registered \uCT\ scan of the same sample. 
Finally, quantitative metrics of Trabecular Thickness
(Tb.Th.), Trabecular Spacing (Tb.Sp.), and Bone Volume to Total Volume (BV/TV) were calculated and evaluated for each approach~\cite{ding_quantification_2000, hildebrand_new_1997, Bouxsein2010}.

\section{Methods}

\begin{table}
\caption{Summary of Notation}\label{tab:symbols}
\begin{tabu}{c X[cm] c}
	\toprule
Variable & {\centering Description} & Nominal Value \\
 & & or Size \\
\midrule
	\textcolor{diffcolor}{$n_\mu$} & \textcolor{diffcolor}{Number of voxels} & --- \\
	\textcolor{diffcolor}{$n_y$} & \textcolor{diffcolor}{Number of measurements} & --- \\
	$\vect{y}$ & Measurements vector & \textcolor{diffcolor}{$n_y$} \\
	$\mat{B}$ & Gain/blur matrix & \textcolor{diffcolor}{$n_y \times n_y$} \\
	$\mat{A}$ & System matrix & \textcolor{diffcolor}{$n_y \times n_\mu$} \\
	$\vect{\mu}$ & Attenuation values vector & \textcolor{diffcolor}{$n_\mu$} \\
	$\mat{K}$ & Measurement covariance & \textcolor{diffcolor}{$n_y \times n_y$} \\
	$\mat{W}$ & Weighting matrix ($\mat{K}^{-1}$) & \textcolor{diffcolor}{$n_y \times n_y$} \\
    $\mR{}$ & Regul\color{diffcolortwo}ar\color{black}izer/penalty function & \textcolor{diffcolor}{$\mathbb{R}^{n_\mu} \to \mathbb{R}$} \\
 $\beta$ & Penalty strength & --- \\
\bottomrule
\end{tabu}
\end{table}

\begin{algorithm*}
\hspace{0.03\textwidth}\begin{tabular}{@{}>{\centering\arraybackslash}p{0.45\textwidth}@{}p{0.02\textwidth}@{}|@{}p{0.02\textwidth}@{}>{\centering\arraybackslash}p{0.45\textwidth}@{}}
Common Initialization
& \multicolumn{2}{@{}>{\centering\arraybackslash}p{0.04\textwidth}@{}}{AND} &
Nesterov Initialization
\\
\begin{algorithmic}
    \State $\vect{\eta} \gets \mat{B}^T \mat{W} \mat{B} \vect{1}$
    \State $\vect{\gamma} \gets \mat{A} \vect{1}$
    \State Calculate $\mat{B}^T \mat{W} \vect{y}$
\end{algorithmic} & & &
\begin{algorithmic}
        \State $\vect{z} \gets \vect{\mu}^{(0)}$
        \State $\vect{w} \gets 0$
        \State $t^{(0)} \gets 1$
        \State $t_{sum} \gets t^{(0)}$
\end{algorithmic}\\
\hline
\multicolumn{4}{@{}p{0.94\textwidth}@{}}{
\begin{algorithmic}
    \For{$p \gets 0..P-1$}
        \For {$m \gets 0..M-1$}
			\State \begin{DIFmanualadd}$n \gets p + m/M, \quad \vect{l}^{(n)} \gets \mat{A}_{(m)} \vect{\mu}^{(n)}, \quad \vect{x}^{(n)} \gets \mathrm{e}^{-\vect{l}^{(n)}}$\end{DIFmanualadd}
            \State \begin{DIFmanualadd}$\vect{\rho}^{(n)} \gets [\mat{B}^T \mat{W} \mat{B}]_{(m)} \vect{x}^{(n)}-[\mat{B}^T \mat{W} \vect{y}]_{(m)} - \vect{\eta}_{(m)} \circ \vect{x}^{(n)} $\end{DIFmanualadd}
            \State \begin{DIFmanualadd}$\vect{L}^{(n)} \gets M \mat{A}^T_{(m)} [-\vect{\eta}_{(m)} \circ \vect{x}^{(n)} \circ \vect{x}^{(n)} - \vect{\rho}^{(n)} \circ \vect{x}^{(n)}]$\end{DIFmanualadd}
			\State \begin{DIFmanualadd} $c^{(n)}_i = \begin{cases} \left[ 2 \frac{\displaystyle 0.5 \eta_i + \rho^{(n)}_i - 0.5 \eta_i (x^{(n)}_i)^2 - x^{(n)}_i \rho^{(n)}_i - l^{(n)}_i \left( \eta_i (x^{(n)}_i)^2 + \rho_i^{(n)} x_i^{(n)} \right) }{\displaystyle (l_i^{(n)})^2} \right]_+ & l^{(n)}_i > 0 \\
			{\left[ 2 \eta_i + \rho^{(n)}_i \right]_+} & l^{(n)}_i = 0 \end{cases}$ \end{DIFmanualadd}
			\State \begin{DIFmanualadd}$\vect{D}^{(n)} \gets M \mat{A}_{(m)}^T \left( \vect{\gamma}_{(m)} \circ \vect{c}^{(n)} \right)$\end{DIFmanualadd}
			\State \begin{DIFmanualadd}Calculate penalty surrogate gradient ($\bigtriangledown \Phi^{(n)}$) and curvature ($\bigtriangledown^2 \Phi^{(n)}$)\end{DIFmanualadd}
            \State $\Delta \vect{\mu} \gets \frac{\displaystyle \vect{L}^{(n)} + \bigtriangledown \Phi^{(n)}}{\displaystyle \vect{D}^{(n)} + \bigtriangledown^2 \Phi^{(n)}}$
\algstore{alg1chkpt1}
\end{algorithmic}
}
\\
\hline
Normal Update
&
\multicolumn{2}{@{}>{\centering\arraybackslash}p{0.04\textwidth}@{}}{OR}
&
Nesterov Update
\\
\begin{algorithmic}
\algrestore{alg1chkpt1}
\State $\vect{\mu}^{(n + 1 / M)} \gets [\vect{\mu}^{(n)} - \Delta \vect{\mu}]_+$
\algstore{alg1chkpt2}
\end{algorithmic}
& & &
                \begin{tabular}[t]{@{}l l@{}}
                \multicolumn{2}{@{}l}{$t^{(n + 1/M)} \gets \frac{1}{2}(1
                +\sqrt{1 + 4 (t^{(n)})^2})$,} \\
                                      $t_{sum} \gets t_{sum} + t^{(n +
                                      1/M)}$, &
                                      $\vect{z} \gets [\vect{\mu}^{(n)} - \Delta \vect{\mu}]_+$, \\
                                      $\vect{w} \gets \vect{w} + t^{(n)} \Delta \vect{\mu}$, &
                                      $\vect{v} \gets [\vect{\mu}^{(0)} - \vect{w}]_+$, \\
                                      \multicolumn{2}{@{}l}{$\vect{\mu}^{(n + 1/M)} \gets \vect{z}
                                      + t^{(n + 1/M)} /t_{sum} (\vect{v} - \vect{z})$}
                 \end{tabular}
\\
\hline
\multicolumn{4}{@{}p{0.94\textwidth}@{}}{
\begin{algorithmic}
\algrestore{alg1chkpt2}
        \EndFor
    \EndFor
\end{algorithmic}
}
\end{tabular}\hspace{0.03\textwidth}
\caption[table]{Algorithm to minimize~\eqref{eq:objective}. The number of
iterations is given by $P$ and the number of ordered subsets by $M$. The
combined iteration and subset index is given by a fractional value,
$n$. For Nesterov acceleration, use \emph{both} initialization columns and
the \emph{right} update column. Otherwise, use the \emph{left}
initialization column and the \emph{left} update column. \begin{DIFmanualadd}Element-wise multiplication is denoted by the $\circ$ operator.
The $[\cdot]_+$ operator returns the (element-wise) maximum of its argument and $0$.\end{DIFmanualadd}}\label{alg}
\end{algorithm*}

\subsection{Forward Model and Objective Function}

A general high-fidelity forward model of the measurements
may be expressed as
\begin{DIFmanualadd}
\begin{equation}
	\bar{y}_i = \sum_k^{n_y} B_{ik} \exp{\left( -\sum_j^{n_\mu} A_{kj} \mu_j \right)}\label{eq:forwardsums}
\end{equation}
where $\bar{y}_i$ is the expected value of measurement $i$ and $\mu_j$ is the attenuation value of voxel $j$.
We assume that the measur\color{diffcolortwo}e\color{black}ments are a sample of a multivariate Gaussian distribution with means given by \eqref{eq:forwardsums} and covariance matrix $\mat{K}$.
By representing the mean measurements as vector $\bar{\vect{y}}$, the attenuation values as vector $\vect{\mu}$,
	and the $B$ and $A$ terms as elements of matrices $\mat{B}$ and $\mat{A}$, respectively, the notation can be
simplified as:
\end{DIFmanualadd}
\begin{subequations}
        \begin{gather}
            \bar{\vect{y}} = \mat{B} \mx{}\label{eq:forwardmeanmodel}\\
            \vect{y} \sim \mathcal{N}(\bar{\vect{y}}, \mat{K})\label{eq:forwardnoisemodel}
        \end{gather}\label{eq:forwardmodel}%
\end{subequations}
where~\eqref{eq:forwardmeanmodel} is the mean forward model and~\eqref{eq:forwardnoisemodel} is the noise model. ($\mathcal{N}$ indicates a normal
distribution, in this case with a mean of $\bar{\vect{y}}$ and covariance matrix
$\mat{K}$.)
\begin{DIFmanualadd}With this notation the exponential function is applied
element-wise. Throughout, matrix and vector variables are boldfaced, and
variables indicating elements of these matrices and vectors are not bold, and
have a subscript indicating which element they refer to (e.g., $y_i$ is the
$i$\textsuperscript{th} element of $\vect{y}$ and $B_{ik}$ is the element at the
$i$\textsuperscript{th} row and $k$\textsuperscript{th} column of
$\mat{B}$).\end{DIFmanualadd} Traditionally, $\mat{A}$ is the forward
projector, $\mat{B}$ is a
diagonal matrix that scales raw transmission values based on the photon fluence
associated with each measurement, and $\mat{K}$ is a diagonal matrix of the
measurement variances (i.e., the covariance matrix with an independent
noise model). While these are conventional choices in the forward model,~\eqref{eq:forwardmodel} is sufficiently general to accommodate more complex
physical characteristics including system blurs (if $\mat{B}$ is a blurring
matrix) and correlated noise (if $\mat{K}$ is a non-diagonal
covariance matrix). In this work, we focus on modeling scintillator and
focal-spot blur as part of $\mat{B}$, and noise correlation due to the scintillator
blur in $\mat{K}$. \begin{DIFmanualadd}Specifically, both blurs are modeled as shift-invariant convolutions.
While this ignores focal-spot geometric effects (e.g., depth dependence) it is an appropriate approximation in
many scenarios (e.g., thin objects, narrow fan angle). See Appendix~\ref{sec:convapprox} in the Supplementary Materials\footnote{available in the supplementary files/multimedia tab} for a derivation of the approximation. (Note that this approximation is not imposed by the forward model,
which is capable of modeling shift-variance and depth dependence.)\end{DIFmanualadd}

Equation~\ref{eq:forwardmodel} also assumes the measurements have an underlying Gaussian
distribution. The PL objective function resulting from~\eqref{eq:forwardmodel} is therefore the penalized non-linear least-squares equation
\begin{equation}
\psi = \frac{1}{2} {\left ( \vect{y} - \mat{B} \mx{} \right )}^T \mat{W} \left ( \vect{y} - \mat{B} \mx{} \right ) + \beta \mR(\vect{\mu}),
\label{eq:objective}
\end{equation}
where $\mR$ is a penalty function \begin{DIFmanualadd}which returns a scalar\end{DIFmanualadd} and $\beta$ is the penalty strength.
The weighting matrix $\mat{W}$ in~\eqref{eq:objective} is the inverse of the covariance matrix
$\mat{K}$ in the forward model~\eqref{eq:forwardmodel}.
The objective function~\eqref{eq:objective} is equivalent (within an additive constant) to
\begin{equation}
\psi_2 = \theta + \beta \mR(\vect{\mu}),
\label{eq:objective2}
\end{equation}
where
\begin{equation}
    \theta \triangleq \frac{1}{2} [\mx{}]^T \mat{B}^T \mat{W} \mat{B} \mx{} -\\
        \vect{y}^T \mat{W} \mat{B} \mx{}.\label{eq:theta}
\end{equation}
The PL reconstruction using this model~\eqref{eq:forwardmodel} is formed by
finding the volume, $\vect{\mu}$, that minimizes
the objective~\eqref{eq:objective2}.
Note that~\eqref{eq:objective2} is non-convex.

We derive
an algorithm to optimize~\eqref{eq:objective2} in a manner similar to that of~\cite{erdogan1999a}, i.e., minimizing a separable quadratic surrogate of the objective
function at each iteration.
Each surrogate matches the objective function in value and first derivative at an operating point, and otherwise majorizes the objective function.
Such an optimization approach is desirable since separability
permits a high degree of parallelization (e.g., facilitating implementation on high-performance GPUs),
while the surrogates framework can guarantee monotonicity.
However, there is a classic trade-off between parallel algorithms, which require many fast iterations, and sequential algorithms, which require fewer slow iterations.
We opt for a separable/parallel algorithm as opposed to a sequential algorithm
to avoid line searches and to take advantage of available GPU hardware.
See~\cite{TilleyII2016} for a steepest-descent algorithm with the same objective function.
In~\cite{erdogan1999a}, separable surrogate
functions are found for the data fit term and the penalty term (in this work
given by $\theta$ and $\mR$, respectively). We use the same formulation for the
penalty term surrogate, but require a new formulation for the data fit term
surrogate. A series of surrogates are calculated for~\eqref{eq:theta}: $Q$,
$Q_2$, and $Q_3$. $Q$ is a surrogate to $\theta$ which is separable in an
intermediate term, $Q_2$ is a quadratic surrogate to $Q$, and $Q_3$ is a
surrogate to $Q_2$ which is both separable in $\vect{\mu}$ and quadratic, and can thus
be easily minimized. Each surrogate function has the same function value and
first derivative as $\theta$ at the current iterate
$(\vect{\mu}^{(n)})$. Therefore, minimizing the final surrogate at every iteration will
monotonically decrease $\theta$~\cite{Erdogan1999}.
The derivation can be found in Appendix~\ref{sec:opttransfer} \begin{DIFmanualadd} (Supplementary Materials\footnote{available in the supplementary files/multimedia tab}),
and the result (\nlpwls{}) is summarized in Algorithm~\ref{alg}.
\end{DIFmanualadd}

Additional
modifications to this underlying update are also shown in Algorithm~\ref{alg}. Specifically, the
algorithm uses the ordered-subsets approach~\cite{erdogan1999a} to accelerate estimation.
The variable $m$ denotes the subset index and subscripts in parentheses indicate that the argument is
modified for the corresponding subset (e.g.\ $\mat{A}_{(m)} \vect{\mu}$ is a forward
projection of $\vect{\mu}$ using only the projection angles in the
$m$\textsuperscript{th} subset). While $p$ indexes the outer loop of iterations
using all of the data, the current iterate $n$ is permitted to take on fractional values indicating progress through
the inner loop of ordered-subsets updates. Algorithm~\ref{alg} also includes a second column of calculations
to optionally apply Nesterov acceleration~\cite{Nesterov2005, Kim2015} to further improve the
rate of convergence.
Note that using ordered subsets or acceleration results in an algorithm that might not converge
(acceleration is only guaranteed to preserve convergence when the objective function is convex).
However, in practice the number of subsets can be chosen such that updates are well-behaved.
Additionally, ordered subsets and acceleration can be used to get close to the solution,
followed by several iterations without subsets or acceleration.
Computationally, each iteration requires one forward projection,
two back projections, and one application of $\mat{B}^T \mat{W} \mat{B}$.

\subsection{Additional Implementation Details}

We apply the proposed algorithm using a model of focal-spot blur and scintillator
blur, where the latter adds spatial correlation to the noise. Both of these
blurs can be represented as factors of the $\mat{B}$ matrix:
\begin{equation}
    \mat{B} \triangleq \mat{B}_\mathrm{d} \mat{B}_\mathrm{s} \mat{G}
    \label{eq:B}
\end{equation}
where $\mat{B}_\mathrm{s}$ is focal-spot blur, $\mat{B}_\mathrm{d}$ is scintillator blur, and $\mat{G}$
scales the data by the bare-beam photon flux per pixel.

As discussed in~\cite{TilleyII2015a}, the covariance matrix of the measurements can be modeled as
\begin{equation}
    \mat{K} = [\mat{B}_\mathrm{d} \mD{}\{\vect{y}\} \mat{B}_\mathrm{d}^T + \mD{}\{\vect{\sigma}_{ro}^2\}]
\end{equation}
where $\vect{\sigma}_{ro}$ is the standard deviation of the readout noise and $\mD{}\{\cdot\}$ is
a diagonal matrix with its argument on the diagonal. 
The weighting matrix $\mat{W}$
is equal to the inverse of $\mat{K}$, which is typically impractical to calculate explicitly.
One approach~\cite{TilleyII2015a} is to use an iterative linear solver to apply the inverse of $\mat{K}$ to the required
vector argument. However, the vector argument is not constant, requiring that the iterative solution be performed every
iteration, substantially increasing
the runtime of the algorithm.
However, note that within the iterative section of the algorithm $\mat{W}$ only appears in
the term $\mat{B}^T \mat{W}
\mat{B}$. One can make the following approximation:
\begin{align}
    \mat{B}^T \mat{W} \mat{B} &= \mat{G}^T \mat{B}_\mathrm{s}^T \mat{B}_\mathrm{d}^T \bigl[ \mat{B}_\mathrm{d} \mD{}\{\vect{y}\} \mat{B}_\mathrm{d}^T \notag \\
    &\quad \quad + \mD{}\{\sigma_{ro}^2\} \bigr]^{-1} \mat{B}_\mathrm{d} \mat{B}_\mathrm{s} \mat{G}\\
    &\approx \mat{G}^T \mat{B}_\mathrm{s}^T \mat{B}_\mathrm{d}^T [ \mat{B}_\mathrm{d} \mD{}\{\vect{y}\} \mat{B}_\mathrm{d}^T ]^{-1} \mat{B}_\mathrm{d} \mat{B}_\mathrm{s} \mat{G}\\
    &= \mat{G}^T \mat{B}_\mathrm{s}^T \mD{}\{1/\vect{y}\} \mat{B}_\mathrm{s} \mat{G}.\label{eq:BTWB}
\end{align}
Equation~\ref{eq:BTWB} can be applied directly \begin{DIFtwo}for\end{DIFtwo} each iteration, and is accurate
when
readout noise is \begin{DIFmanualadd}small relative to the measurements\end{DIFmanualadd}, $\mat{B}_\mathrm{d}$ is invertible, and all measurements
are greater than 0.
Note that $\mat{B}^T \mat{W} \mat{B}$ is only the weighting term, so removing $\mat{B}_\mathrm{d}$ from this term is not equivalent to removing scintillator blur from the model.
Scintillator blur is still included in other applications of $\mat{B}$ (see next paragraph).
Additionally,~\eqref{eq:BTWB} ensures $\vect{\eta}$ (see Algorithm~\ref{alg}) is
always positive (a requirement of the optimum curvature derivation, see Appendix~\ref{sec:optc}\begin{DIFmanualadd}, Supplementary Materials\end{DIFmanualadd}\footnote{available in the supplementary files/multimedia tab}) as long as $\mat{B}_\mathrm{s}$ and $\mat{G}$ are non-negative, all diagonal
elements of $\mat{G}$ are positive, and $\mat{B}_\mathrm{s}$ has no rows or columns of all
zeros.

The inverse covariance matrix is also included in the term $\mat{B}^T \mat{W} \vect{y}$,
which appears in the initialization section of the algorithm.
In this work \num{200} iterations of a preconditioned conjugate gradient method~\cite{nocedal_numerical_2006, hestenes1952methods} were used to calculate
this term. The preconditioning matrix was $\mD{}\{\vect{y} + \vect{\sigma}^{\begin{DIFmanualadd}2\end{DIFmanualadd}}_{ro}\}$.

\subsection{System Characterization}
\label{methods:mtf}

To evaluate the proposed reconstruction method, scintillator and focal-spot blur properties of a prototype
extremities qCBCT test bench~\cite{marinetto_quantification_2016} were first characterized.
This characterization was then used to ensure an accurate simulation study (\S\ref{simulation}) and to generate
accurate blur models for \nlpwls\ reconstructions of physical test-bench data
(\S\ref{realdata}). The test bench uses an IMD RTM37 rotating anode X-ray source (with dual 0.3/0.6
focal spots) and a Teledyne DALSA Xineos3030HR CMOS X-ray detector
(\SI{100}{\micro\meter} pixel pitch and CsI scintillator). The geometry emulates
that of a prototype extremities qCBCT system, with a source-to-detector
distance of \SI{51}{\centi\meter} and a source-to-axis distance of
\SI{38}{\centi\meter}. X-ray focal-spot and
detector blur were estimated from a pinhole image of the focal spot, edge spread
function (ESF) measurements at the detector (where focal-spot blur is
negligible), and ESF measurements at isocenter. The readout noise
($\sigma_{ro}$) was estimated using dark scans.

Images of a tungsten edge were used to calculate ESFs, which in turn were used
to calculate modulation transfer functions (MTFs) as described
in~\cite{TilleyII2015a, Samei1998}. MTFs were measured in two directions along the detector: parallel
to the axes of rotation (axial) and perpendicular to the axis of rotation
(trans-axial).
This work assumes the detector scintillator MTF is radially symmetric and uses the model
of~\cite{siewerdsen_signal_1998} with an additional Gaussian component to capture observed low frequency characteristics:
\begin{equation}
    MTF_{d} = g \mathrm{e}^{-f^2 / \sigma^2} + (1 - g) (1 + H f^2)^{-1}
    \label{eq:scintmtf}
\end{equation}
where $f$ is frequency and $g$ is the relative strength of the
Gaussian term (between 0 and 1). Combined with pixel sampling, the MTF model
at the detector is
\begin{equation}
   MTF_{da} = \sinc(f T) MTF_{d}  
   \label{eq:detmtf}
\end{equation}
where $T$ is the pixel pitch. Because the pixels are square and the scintillator
MTF is assumed to be radially symmetric,~\eqref{eq:detmtf} models both the
horizontal and vertical MTFs. We estimated the parameters $g$, $\sigma$, and $H$
by fitting~\eqref{eq:detmtf} to the MTFs measured at the detector.

Pinhole images of the X-ray focal spot were acquired using a 07--633 pinhole
assembly (Fluke Electronics, Everett, WA) with a nominal diameter of
\SI{0.010}{\milli\meter}. A
point spread function (PSF) that
models the focal-spot blur experienced by an object at isocenter was found using
this pinhole image. Because the
pinhole was imaged at a high magnification (\num{\sim34}), multiple
manipulations were required to obtain the final PSF. First, scale factors were
found for each axes to match the shape of the pinhole image to that of the focal-spot
PSF at isocenter.
We chose the scaling parameters by fitting the axial and
trans-axial slices of the pinhole
derived MTF to the MTFs measured with the tungsten edge at isocenter.
The
axial and trans-axial scaling parameters are not necessarily the same due to
different shift-variant properties in these two directions, and the possibility that
the pinhole was slightly misaligned. The pinhole image was resampled using these
scaling parameters to produce a super-sampled PSF of the focal-spot blur at
isocenter. In order to account for the aperture of each pixel, the super-sampled
PSF was convolved with a 
\SI{100x100}{\micro\meter} rect function corresponding to the
pixel pitch and then binned and normalized to produce a PSF with
\SI{100}{\micro\meter} pixels (i.e., in native measurement dimensions).

\subsection{Simulation Study}
\label{simulation}

\begin{figure}
\centering
\leavevmode\beginpgfgraphicnamed{tille1}%
\input{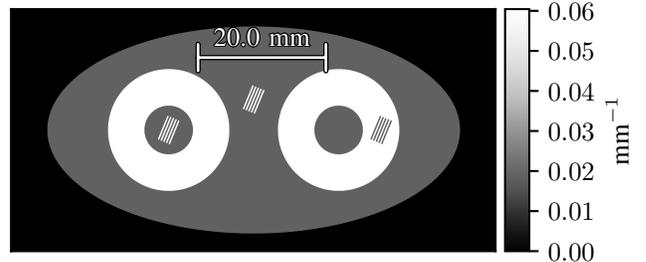}
\endpgfgraphicnamed%
\caption[figure]{\begin{DIFmanualadd}Digital phantom with line pairs and bone inserts. The background attenuation value in the oval
	is \SIr{3}{\fat}{\per\milli\meter} and the bone attenuation is
	\SIr{3}{\bone}{\per\milli\meter}. The line pair attenuation values are
	either \SIr{3}{\bone}{\per\milli\meter} (left and center) or
	\SIr{3}{\fat}{\per\milli\meter} (right).\end{DIFmanualadd}
	\begin{DIFtwo}The line pair frequency is \SI{2.38}{\per\milli\meter}.\end{DIFtwo}}\label{fig:lpphantom}  
\end{figure}

Data were generated using the digital phantom in
Fig.~\ref{fig:lpphantom} and a simulated system model based on the test bench geometry and characterization.
Specifically,
the high-resolution phantom was created with
\begin{DIFmanualadd}\SI{17.5x17.5x70}{\micro\meter} voxels (with the long axis of the voxel parallel to the axis of rotation)\end{DIFmanualadd} and high contrast line pairs with an attenuation of \SIr{3}{\bone{}}{\per\milli\meter} (bone) and a background
attenuation of \SIr{3}{\fat}{\per\milli\meter} (fat). To model nonlinear partial volume effects, this 
phantom was forward projected onto a
\SI{87.5}{\milli\meter} detector \begin{DIFmanualadd}of subpixels\end{DIFmanualadd} with a small pixel pitch
\begin{DIFmanualadd}(\SI{25x100}{\micro\meter})\end{DIFmanualadd} at 720 equally spaced angles
using a separable footprints model~\cite{Long2010} for the projector. 
\begin{DIFmanualadd}The forward model for data generation used finite integration over the extended focal spot and detector elements:
\begin{equation}
	\vect{y} = \mat{S} \tilde{\mat{B}}_\mathrm{d} \tilde{\mat{G}} \sum_k^{n_k} \omega_k e^{-\mat{A}_k \mu}
\end{equation}
where $\mat{A}_k$ is a projection matrix corresponding to an individual sourcelet with relative intensity $\omega_k$,
$\tilde{\mat{B}}_\mathrm{d}$ is a scintillator blur~\eqref{eq:scintmtf} matrix which operators on subpixels, $\tilde{\mat{G}}$ scales the subpixels by the photon flux, and $\mat{S}$ bins the subpixels to \SI{100x100}{\micro\meter} pixels.
To obtain a final photon flux of \SI{e3}{\photon\per\pixel}, $\tilde{\mat{G}}$ scaled each subpixel by \SI{250}{\photon\per\pixel}.
The
scintillator blur matrix was applied functionally using
Fourier operations and nearest neighbor substitution at the
boundaries. 
Focal-spot blur was modeled by forward projecting with 354 sourcelets derived from the super-sampled PSF from \S\ref{methods:mtf} (summed to one dimension).
The modeled anode angle was \ang{17.5}.\end{DIFmanualadd}
Noisy data were generated from a Poisson distribution with the
Poisson parameter equal to the pre-scintillator-blur data (e.g., the vector
\begin{DIFmanualadd}before\end{DIFmanualadd}
application of \begin{DIFmanualadd}$\tilde{\mat{B}}_\mathrm{d}$\end{DIFmanualadd}), and these noisy data were
blurred by \begin{DIFmanualadd}$\tilde{\mat{B}}_\mathrm{d}$\end{DIFmanualadd}. Finally, we added Gaussian readout noise with a standard
deviation of \SI{7.109}{\photon} (based on bench data dark scan values) to
obtain the final measurements.

In all simulation studies the reconstruction volume was \begin{DIFmanualadd}\SI{70x35}{\milli\meter}\end{DIFmanualadd} with \SI{0.07}{\milli\meter}
cubic voxels (i.e.,~approximately equal to the demagnified pixel size).
Data were reconstructed with the presented \nlpwls{} algorithm incorporating \begin{DIFmanualadd} the blur models derived in \S\ref{methods:mtf}.
	Specifically, $\mat{B}$ in~\eqref{eq:B} was applied, where $\mat{B}_\mathrm{s}$ and $\mat{B}_\mathrm{d}$ convolve their inputs with the focal-spot PSF (summed to one dimension) and
the scintillator blur~\eqref{eq:scintmtf}, respectively, and $\mat{G}$ scales each value by \SI{e3}{\photon\per\pixel}.
With the low photon flux of the simulation study, the measurement data is not substantially higher than readout noise, and \eqref{eq:BTWB} is not
a valid approximation. Therefore, \num{20} iterations of the preconditioned conjugate gradient method were used to apply $\mat{W}$ every iteration.\end{DIFmanualadd}
For comparison, the same optimization strategy was used with two different forward models.
The first, \nlpwlsnc{}, assumed the noise was uncorrelated (i.e., $\mat{K} = \mD{}\{\vect{y} + \vect{\sigma}^{\begin{DIFmanualadd}2\end{DIFmanualadd}}_{ro}\}$).
The second, \nlpwlsi{}, also assumed the noise was uncorrelated, and additionally assumed there was no blur (i.e., $\mat{B} = \mat{G}$).
Finally, \begin{DIFmanualadd}the data were also reconstructed using Fourier domain deblurring (using the same blur models as \nlpwlsnc{} and \nlpwls{}) followed by FDK (\fdkdeblur{}) with multiple cutoff frequencies.\end{DIFmanualadd}
All model-based reconstructions used the separable footprints projector~\cite{Long2010}.

Reconstructions were assessed with \begin{DIFmanualadd}three\end{DIFmanualadd} metrics: bias, noise, and maximum Jaccard index (mJac)~\cite{jaccard_distribution_1912}.
\begin{DIFtwo}Bias and noise were chosen as traditional image quality metrics, while mJac was picked as a metric specific to trabecular bone analysis.\end{DIFtwo}
These metrics were calculated for \begin{DIFmanualadd}the set of line pairs in the middle of Figure~\ref{fig:lpphantom}\end{DIFmanualadd}.
The terms are defined based on the truth image $\vect{t}$ (binned to match voxel size),
reconstructions of noiseless data $\vect{\hat{\mu}}_{n\color{diffcolortwo}l}(\beta, \delta)$, reconstructions of noisy data $\vect{\hat{\mu}}(\beta, \delta)$, \begin{DIFtwo}and the number of voxels in the ROI ($N_{roi}$)\end{DIFtwo}
\begin{DIFmanualadd}\begin{gather} %
	\bias(\beta, \delta) = \|\vect{\hat{\mu}}_{n\color{diffcolortwo}l}(\beta, \delta) - \vect{t} \| \color{diffcolortwo} / N_{roi} \color{black}\\
	\noise(\beta, \delta) = \| \vect{\hat{\mu}}(\beta, \delta) -  \vect{\hat{\mu}}_{n\color{diffcolortwo}l}(\beta, \delta) \| \color{diffcolortwo} / N_{roi}. \color{black}
\end{gather}
These metrics were calculated in an ROI encompassing the central line pairs.\end{DIFmanualadd}
To calculate mJac, a truth segmentation $\vect{t}_b$ was calculated by thresholding the truth image $\vect{t}$ at \SIr{3}{\avgfatbone}{\per\milli\meter} (the average attenuation value of fat and bone).
The reconstruction $\vect{\hat{\mu}}$ was thresholded by a value $t$ for 101 values of $t$ between the attenuation values of fat and bone, inclusive.
The mJac value for a given reconstruction is the maximum Jaccard index between the truth segmentation and the segmented $\vect{\hat{\mu}}$ over all $t$:
\begin{DIFmanualadd}\begin{equation}
    \mJac(\beta, \delta) =\max_t \left[ \jaccard \left( \vect{\hat{\mu}}( \beta, \delta ) > t, \vect{t}_b \right) \right] \color{diffcolortwo}.\color{black}\label{eq:mJac}
\end{equation}\end{DIFmanualadd}
The Jaccard index ranges from 0 to 1, with 1 indicating perfect
correspondence with the truth segmentation. 

\subsubsection{Parameter Sweep}

This work used a Huber penalty for the regularizer (\begin{DIFmanualadd}$\mR$\end{DIFmanualadd})~\cite{huber_robust_statistics}.
The Huber penalty has an additional parameter, $\delta$, which is the value below which pixel differences will be penalized quadratically.
We conducted a parameter sweep over $\beta$ and $\delta$ in order to pick an appropriate value for $\delta$.
Phantom data were reconstructed using \nlpwls{} and \begin{DIFmanualadd}\nlpwlsi{}.
Additionally, two photon fluxes were used: \SI{e3}{\photon\per\pixel} (low photon flux) to match the simulation study, and \SI{4e4}{\photon\per\pixel} (high photon flux) to approximate the bench study.
The high photon flux data utilized the covariance matrix approximation in~\eqref{eq:BTWB}.\end{DIFmanualadd}
Both algorithms used \begin{DIFmanualadd}501\end{DIFmanualadd} iterations, 10 subsets,
and momentum-based acceleration.
\begin{DIFmanualadd}The mJac metric was calculated for each ($\beta$, $\delta$) pair.\end{DIFmanualadd}

\subsubsection{Algorithm Comparison}

\begin{DIFmanualadd}\fdkdeblur{}\end{DIFmanualadd}, \nlpwlsi{}, \nlpwlsnc{}, and \nlpwls{} were compared by analyzing the bias/noise tradeoff and mJac over a range of regularization strengths.
A large number of iterations (\num{20000}) were used to ensure nearly converged estimates.
We utilized a scheduling approach for acceleration and the number of subsets, with \num{50} iterations of acceleration and \num{10} subsets,
followed by \num{50} iterations of acceleration and \num{5} subsets, \num{10000} iterations of acceleration and no subsets, and finally \num{9000} iterations of
no acceleration and no subsets.
We used a Huber penalty with $\delta =$ \begin{DIFmanualadd}\SI{e-2}{\per\milli\meter}\end{DIFmanualadd}.
A \begin{DIFmanualadd}bias/noise\end{DIFmanualadd} plot and a plot of mJac as a function of $\beta$ were analyzed for \begin{DIFmanualadd}the center set of line pairs in Figure~\ref{fig:lpphantom}
\end{DIFmanualadd}and each of the four reconstruction methods.
For direct visual comparison we present reconstructions of the line pairs, along with the corresponding optimum segmentations.

\subsection{Bench Data}
\label{realdata}

To investigate the performance of the proposed algorithm on physical
data, a human iliac-crest bone-biopsy core was scanned on the test bench described
in \S\ref{methods:mtf}. The bone sample comprised both trabecular and cortical
bone. $\mat{B}$ was modeled as described previously~\eqref{eq:B}, with $\mat{B}_\mathrm{s}$
and $\mat{B}_\mathrm{d}$ representing applications of the models developed in
\S\ref{methods:mtf}. Blur matrices were applied functionally as in the simulation study.
\begin{DIFmanualadd}$\mat{B}_\mathrm{d}$ was applied using Fourier methods and $\mat{B}_\mathrm{s}$ was applied using convolution.
The covariance approximation~\eqref{eq:BTWB} was used.\end{DIFmanualadd}
$\mat{G}$ was a matrix which scaled the values of each pixel by the estimated
bare-beam photon flux and each frame by a normalization factor.\footnote{Details
are given in Appendix~\ref{sec:gain} in the supplementary material, available in the supplementary files/multimedia tab.}
The projection operator $\mat{A}$ used the separable footprints algorithm as in the
simulation study. The GPL methods used the same readout noise value as the simulation study.

Reconstructions were initialized with FDK and ran for 650 iterations with 10
subsets to obtain well-converged estimates. The trabecular bone was also reconstructed with \begin{DIFmanualadd}\nlpwlsi{} and \nlpwlsnc{}\end{DIFmanualadd} using the same number of
iterations and subsets. Momentum-based acceleration was applied in \begin{DIFmanualadd}all cases\end{DIFmanualadd}.
A Huber regularization penalty was used with a range of penalty
strengths and $\delta$ equal to \begin{DIFmanualadd}\SI{1e-3}{\per\milli\meter}\end{DIFmanualadd}~\cite{huber_robust_statistics}. We
also computed an FDK reconstruction (frequency cutoff at Nyquist and no
additional apodization) for comparison.
In all cases the reconstruction volume was \SI{60x60x30}{\milli\meter} with \SI{0.075}{\milli\meter} voxels (i.e.,~voxel size was approximately equal to the demagnified pixel size).
The projection area was \SI{120x25}{\milli\meter} with \SI{0.1}{\milli\meter} pixel pitch and 720 frames.

Reconstructions of qCBCT data were compared with high resolution \uCT{} data using
mJac~\eqref{eq:mJac}, Trabecular Thickness (Tb.Th.), Trabecular Spacing (Tb.Sp.), and Bone Volume to Total Volume fraction (BV/TV)~\cite{ding_quantification_2000, hildebrand_new_1997, Bouxsein2010}.
Bench data were acquired at \SI{90}{\kilo\volt} and \SI{90.7}{\milli\ampere\second}.  %
The \uCT{} data were acquired on a SkyScan 1172 CT scanner (Bruker microCT, Kontich, Belgium) at \SI{65}{\kilo\volt}.
To find the ``true'' trabecular bone segmentation with the same voxel size as the
reconstructions, the \uCT{} image of the trabecular bone was first binned from
\SI{0.0076}{\milli\meter\per\voxel} to
\SI{0.0380}{\milli\meter\per\voxel} and then registered
with an FDK reconstruction of the qCBCT bench data. The registration algorithm also
reduced the voxel size of the \uCT{} image to match that of the FDK reconstruction (and therefore the
\begin{DIFmanualadd}model-based\end{DIFmanualadd} reconstructions). The resulting image is referred to as
\uCT{}mv for Matched Voxel size. The Elastix software package~\cite{klein_elastix:_2010} registered the images using the  %
binned \uCT{} reconstruction as the moving image, a similarity transformation, and the 
Mutual Information Metric. A mask  %
was used to limit the evaluation of the registration metric to a sub-volume
containing only trabecular bone. The \uCT{}mv image was
thresholded to generate the ``truth segmentation.'' The threshold value was
chosen using a visual histogram inspection. The FDK,
\begin{DIFmanualadd}\nlpwlsi{}, \nlpwlsnc{}\end{DIFmanualadd}, and \nlpwls{} reconstructions were thresholded at 101 equally
spaced attenuation values between \SI{0}{\per\milli\meter} and
\SI{0.07}{\per\milli\meter}, inclusive, to calculate mJac.
The mJac metric was only computed within the trabecular region (using the
same mask as the
registration). This metric was plotted for each
MBIR reconstruction method as a function of regularization strength. 
The most accurate segmented reconstruction from each MBIR method was selected 
as the one with the highest mJac over all
regularization strengths, and the most accurate
reconstruction was selected as the corresponding pre-thresholded image. The optimal FDK
segmentation was defined as the one with the highest mJac over all
threshold values.
A Tb.Th.\ map was calculated from the optimal segmented
reconstruction for each reconstruction method and the \uCT{}mv image.
Tb.Th.\ and Tb.Sp.\ were calculated with BoneJ~\cite{doube_bonej:_2010}, a plug-in for ImageJ~\cite{schneider_nih_2012}.
Average Tb.Th., Tb.Sp, and BV/TV were computed over the area defined
by the registration/mJac mask.
The Tb.Th.\ and BV/TV of the original \uCT\ image (before binning and
registration) were also calculated using the same mask (transformed to the
\uCT\ coordinates). (Tb.Sp.\ was not calculated for this image due to computation constraints.) Slices of the \uCT\ scan and \uCT\ Tb.Th.\ map were transformed using the registration parameters calculated previously, facilitating visual comparison to the other methods. 
Optimal reconstructions, optimal segmentations, and Tb.Th.\ maps for
FDK, \begin{DIFmanualadd}\nlpwlsi{}, \nlpwlsnc{}\end{DIFmanualadd}, and \nlpwls{} were compared with corresponding 
\uCT{}mv and original \uCT\ images.

\section{Results}

\subsection{System Characterization}

\begin{figure}
\centering
\leavevmode\beginpgfgraphicnamed{tille2}%
\input{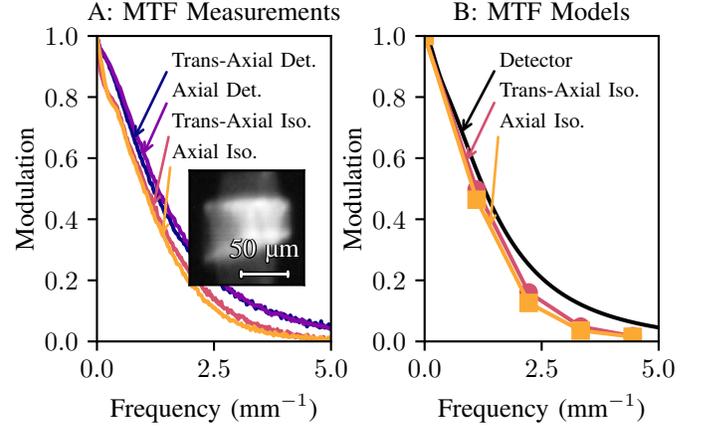}
\endpgfgraphicnamed%
\caption{System Characterization Results. A: Measured axial and
trans-axial MTF slices derived from tungsten
edge responses.
Inset: Pinhole image of the X-ray
focal spot, resampled to match the PSF of the focal-spot blur experienced by an
object at isocenter.
B: MTF models. The detector model has the form
given in~\eqref{eq:detmtf}. The isocenter models are slices of the MTF derived
from the
final PSF multiplied by the detector MTF model.
}\label{fig:mtfs}
\end{figure}

The system characterization results are shown in Fig.~\ref{fig:mtfs}.  The
measured MTFs are plotted in Fig.~\ref{fig:mtfs}A and show that, for the
prototype test bench, detector blur
is a larger effect than focal-spot blur. Because detector blur (scintillator
blur and pixel aperture blur) is the same at the detector and at isocenter, the
difference between the isocenter MTF and the detector MTF is due to focal spot
blurring. This difference is relatively small,  indicating that this system is dominated by
detector blur. The axial and trans-axial detector MTFs are almost equivalent,
supporting the radially symmetric assumption used in the model. The MTF models
(Fig.~\ref{fig:mtfs}B) strongly match the measured data.

The focal-spot pinhole image was scaled and resampled to match the
magnitude of the blur experienced by an object at isocenter
(Fig.~\ref{fig:mtfs}). The focal spot has
a primary trapezoidal component with a higher intensity on two of the edges,
similar to those observed on
the rectangular focal spot studied previously~\cite{TilleyII2015a}.
This focal spot has an additional, lower intensity, trapezoidal
component with a different orientation, creating a cross pattern. Because of
this complicated structure, we decided to derive a PSF directly from the pinhole
image instead of using a mathematical model. The scale bar illustrates that the
focal spot blur is relatively small (about the size of a detector pixel) for an
object at isocenter.

\begin{figure}
\centering
\leavevmode\beginpgfgraphicnamed{tille3}%
\input{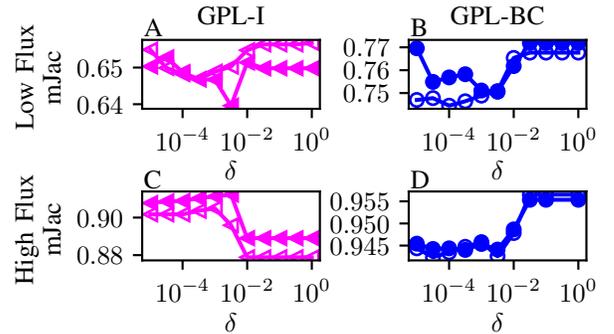}
\endpgfgraphicnamed%
    \caption[figure]{\begin{DIFmanualadd}Parameter sweep results. Each point is the maximum mJac over $\beta$ for a given $\delta$\begin{DIFtwo},\end{DIFtwo} reconstruction method\begin{DIFtwo}, and noise realization\end{DIFtwo}. The left column is \nlpwlsi{} and the right column is \nlpwls{}. The top row is the low photon flux results and the bottom row is the high photon flux results.\end{DIFmanualadd}}\label{fig:betadelta}
\end{figure}

\begin{figure}
\centering
\leavevmode\beginpgfgraphicnamed{tille4}%
\input{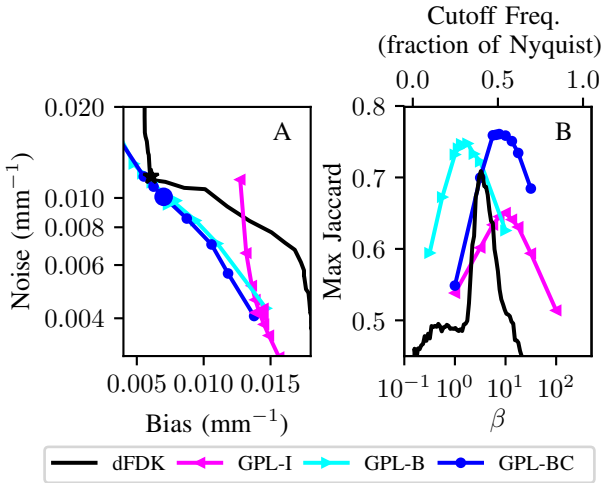}
\endpgfgraphicnamed%
	\caption[figure]{\begin{DIFmanualadd}Bias/noise\end{DIFmanualadd} \begin{DIFmanualadd}(A)\end{DIFmanualadd} and mJac \begin{DIFmanualadd}(B)\end{DIFmanualadd} plots.
		The large markers in \begin{DIFmanualadd}(A)\end{DIFmanualadd} correspond to the maximum mJacs in \begin{DIFmanualadd}(B).
	The frequency cutoffs for the \fdkdeblur{} data in B (x-axis) are indicated at the top of the plot.\end{DIFmanualadd}}\label{fig:sim}
\end{figure}

\begin{figure*}
\centering
\leavevmode\beginpgfgraphicnamed{tille5}%
\input{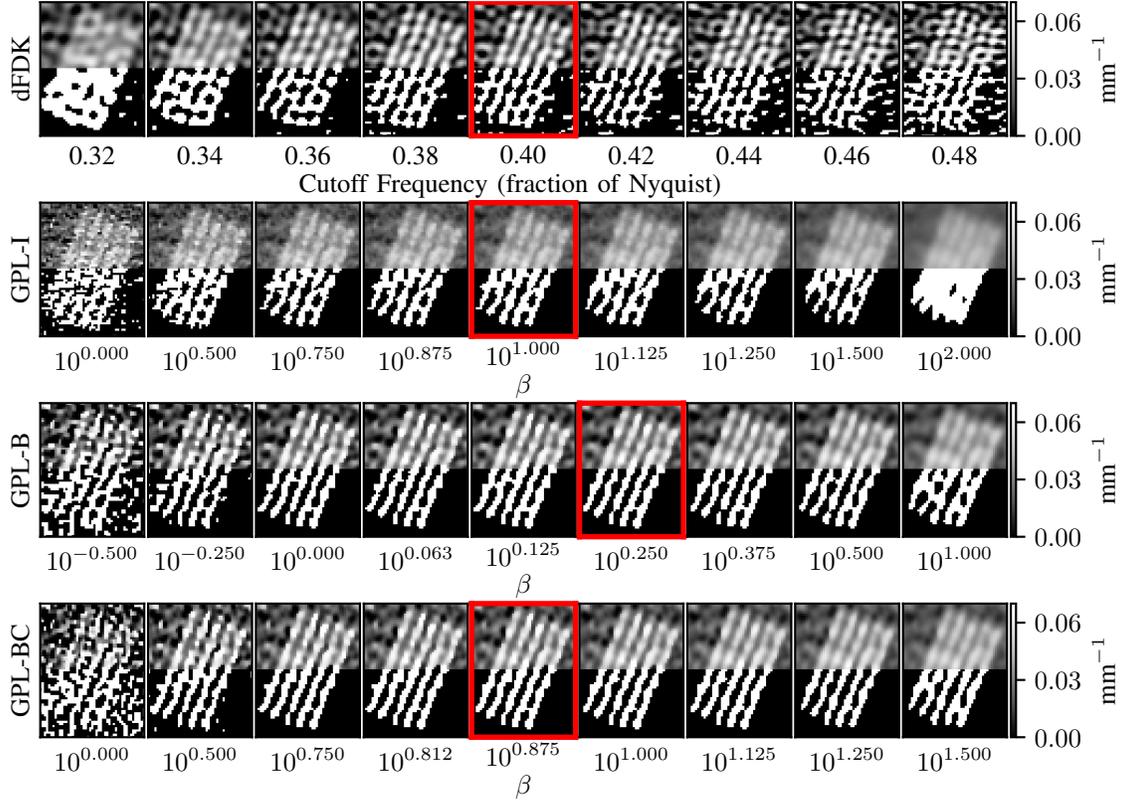}
\endpgfgraphicnamed%
	\caption[figure]{The \begin{DIFmanualadd}center\end{DIFmanualadd} line pairs from the reconstructions in Fig.~\ref{fig:sim}. Each row corresponds to a different reconstruction method. Note that different values of $\beta$ were used for different \begin{DIFmanualadd}MBIR\end{DIFmanualadd} methods. The reconstructions with the red border correspond to the ones with the maximum mJac in Fig.~\ref{fig:sim}\begin{DIFmanualadd}(B)\end{DIFmanualadd}. The lower half of each image shows the best segmentation for that \begin{DIFmanualadd}$\beta$/cutoff\end{DIFmanualadd} (i.e., the one resulting in the maximum Jaccard index over threshold values).}\label{fig:simrecs}
\end{figure*}

\subsection{Simulation Study}

\begin{figure}
\centering
\leavevmode\beginpgfgraphicnamed{tille6}%
\input{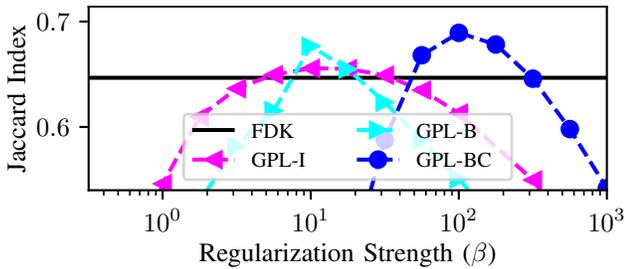}
\endpgfgraphicnamed%
	\caption{Maximum Jaccard (mJac) for each reconstruction method and regularization
strength for the test bench-data.}\label{fig:real_mo}
\end{figure}

\begin{figure}
\centering
\leavevmode\beginpgfgraphicnamed{tille7}%
\input{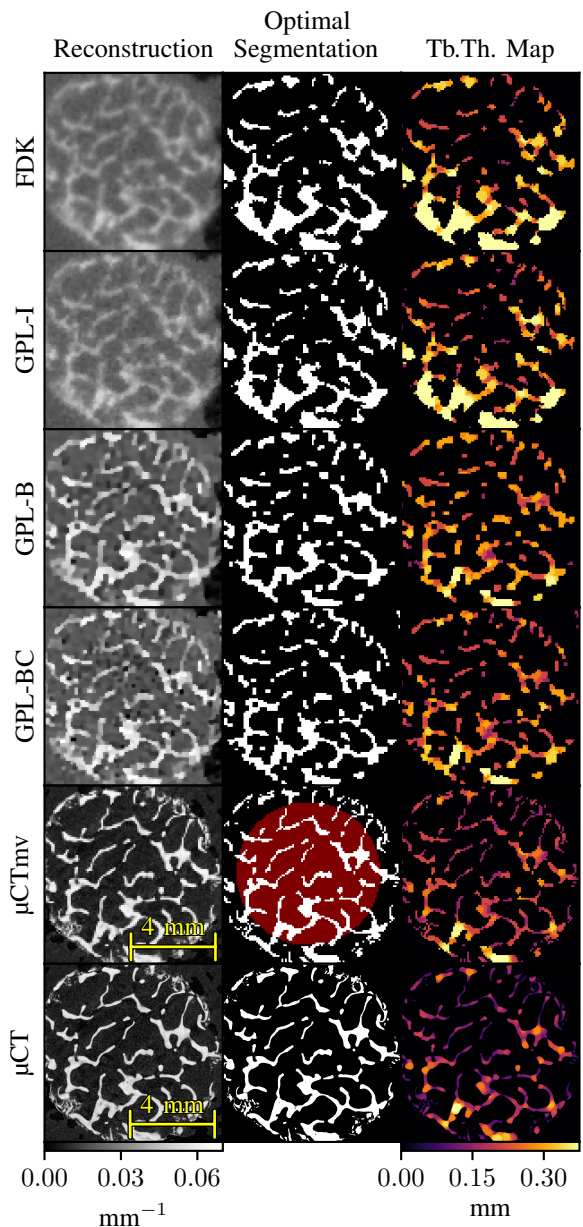}
\endpgfgraphicnamed%
\caption{Axial slice of trabecular bone reconstructions. Rows correspond to
different reconstruction methods. The red background in the \uCT{}mv segmentation
image indicates a slice of the mask used for registration and metric calculation. Note the attenuation values
of the \uCT\ and \uCT{}mv scans are arbitrary and do not correspond to the gray-level map
on the bottom.}\label{fig:real_a}
\end{figure}

\begin{figure}
\centering
\leavevmode\beginpgfgraphicnamed{tille8}%
\input{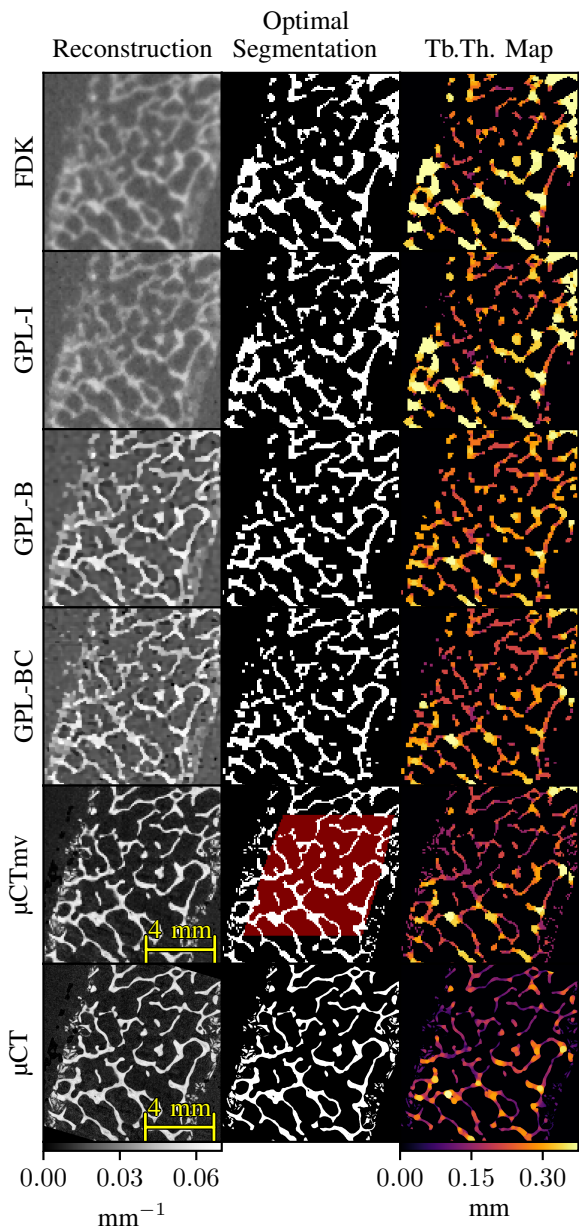}
\endpgfgraphicnamed%
\caption{Trans-axial slice of trabecular bone reconstructions. \uCT\ and
\uCT{}mv attenuation units are arbitrary.}\label{fig:real_c}
\end{figure}

\subsubsection{Parameter Sweep}
\begin{DIFmanualadd}Fig.~\ref{fig:betadelta} shows the maximum mJac over $\beta$ as a function of $\delta$ \begin{DIFtwo} for two noise realizations\end{DIFtwo}.\end{DIFmanualadd}
\begin{DIFmanualadd}These results\end{DIFmanualadd} indicate that mJac \begin{DIFmanualadd}is\end{DIFmanualadd} relatively insensitive to $\delta$ \begin{DIFmanualadd}(compare the ranges in the plots in Fig.~\ref{fig:betadelta} to
those in Fig.~\ref{fig:sim}). \begin{DIFtwo}This is potentially due to the fact that mJac is insensitive to edge smoothness.\end{DIFtwo}
\begin{DIFtwo}The measurements are relatively noisy at this scale, especially with low gain and low $\delta$.\end{DIFtwo}
For \nlpwls{} the optimal $\delta$ is higher than any contrast in the phantom,
indicating a ``near'' quadratic penalty is ideal.
The simulation data were reconstructed with $\delta =$ \SI{e-2}{\per\milli\meter} \begin{DIFtwo}(where mJac values are high and stable)\end{DIFtwo} and the bench data with $\delta =$ \SI{e-3}{\per\milli\meter} (which 
potentially gives a slight advantage to \nlpwlsi{}).\end{DIFmanualadd}

Figure~\ref{fig:sim} shows the \begin{DIFmanualadd}bias/noise\end{DIFmanualadd} trade-off \begin{DIFmanualadd}(A)\end{DIFmanualadd} and maximum Jaccard index \begin{DIFmanualadd}(B)\end{DIFmanualadd} for
\begin{DIFmanualadd}the center set of line pairs\end{DIFmanualadd}. \begin{DIFtwo}Results are similar but less dramatic for the other two sets of line pairs (not shown).\end{DIFtwo}
At lower regularization strengths, reconstructions of noiseless data are more accurate (lower bias),
but reconstructions of noisy data result in noisy reconstructions.
On the other hand, for higher regularization strengths, noise is suppressed at the cost of increased smoothing/blurring
of the image, imparting bias.
Methods with blur modeling (\nlpwls{} and \nlpwlsnc{}) were able to achieve a lower bias than \begin{DIFmanualadd}the method\end{DIFmanualadd} without blur modeling
(\nlpwlsi{}).
\begin{DIFmanualadd}\nlpwls{} and \nlpwlsnc{} have a similar bias/noise trade-off, with \nlpwls{} showing a slight advantage.
Because it does not include a blur model, \nlpwlsi{} encounters a bias limit at about \SI{0.013}{\per\milli\meter}.
\fdkdeblur{} can achieve lower bias reconstructions than \nlpwlsi{}, but suffers from increased noise as compared to
\nlpwlsnc{} and \nlpwls{}. However, there is a small range (near the best \fdkdeblur{} mJac performance) where \fdkdeblur{} performs comparably to \nlpwlsnc{} and \nlpwls{}.\end{DIFmanualadd}

Figure~\ref{fig:sim}\begin{DIFmanualadd}B shows\end{DIFmanualadd} similar trends.
For each method the ``best'' reconstruction is defined as the one with the maximum mJac.
This maximum mJac value is used to compare the different methods.
\begin{DIFmanualadd}\nlpwls{} results in the best reconstruction, followed by \nlpwlsnc{},
\fdkdeblur{}, and \nlpwlsi{}.
The advantage of \nlpwls{} over \nlpwlsnc{} is more apparent in the mJac plot than the bias/noise plot.\end{DIFmanualadd}

Figure~\ref{fig:simrecs} shows reconstructions of the \begin{DIFmanualadd}center line pairs\end{DIFmanualadd}.
The bottom half of each image shows the optimal segmentation (i.e., the one resulting in the best mJac).
\begin{DIFmanualadd}\nlpwlsi{} results in the worst performance with low contrast line pairs.
The line pairs in the \fdkdeblur{} reconstruction are more distinct but both the line pairs and the background exhibit increased noise.
Finally, the \nlpwlsnc{} and \nlpwls{} reconstructions have less noise than the \fdkdeblur{} reconstruction without sacrificing line pair visualization.
The difference between the \nlpwls{} and \nlpwlsnc{} reconstructions is subtle, but can be appreciated in the thresholded image, where the
\nlpwls{} method results in thicker and more uniform line pairs.
The noise difference between \nlpwls{}/\nlpwlsnc{} and \fdkdeblur{} is particularly evident in the background of the segmented image, where the \fdkdeblur{}
reconstruction contains noisy values above the segmentation threshold, resulting in a ``splotchy'' segmented background image.\end{DIFmanualadd}
Qualitatively, the visually ``best'' reconstructions correspond to those with the best mJac (indicated by a red outline),
confirming the suitability of this metric.

\subsection{Bench Data}

This section presents the results of the prototype test-bench study with human
trabecular bone. The mJac for each
reconstruction
is shown as a function of regularization strength in Fig.~\ref{fig:real_mo}.
The \nlpwls{} method is able to achieve \begin{DIFmanualadd}the highest\end{DIFmanualadd} maximum 
mJac\begin{DIFmanualadd}, followed by \nlpwlsnc{}, \nlpwlsi{}, and\end{DIFmanualadd}
FDK (indicated by the black line).
The optimal
segmentation thresholds for the most accurate \begin{DIFmanualadd}\nlpwlsi{}, \nlpwlsnc{},\end{DIFmanualadd} and \nlpwls{} reconstructions (i.e., those with the maximum mJac over regularization strength, corresponding to the maxima in Fig.~\ref{fig:real_mo}) are
\begin{DIFmanualadd}\SI{\nlpwlsspsIthresholdopt}{\per\milli\meter}, \SI{\nlpwlsspsBthreethresholdopt}{\per\milli\meter},\end{DIFmanualadd} and
\SI{\nlpwlsspsthresholdopt}{\per\milli\meter}, respectively.

The most accurate reconstructions
are shown in Fig.~\ref{fig:real_a} and Fig.~\ref{fig:real_c}, along with the
corresponding segmented trabecular bone images (using the optimal thresholds) and
Tb.Th.\ maps. The FDK reconstruction, the registered \uCT\
reconstruction with Matched Voxel size (\uCT{}mv),
and the registered \uCT\ reconstruction slices with the original \uCT\ voxel size (\uCT) are also
included. While the \uCT\
reconstruction is the
best approximation of the true image volume, the \uCT{}mv image is a better
approximation of the best achievable reconstruction at the chosen voxel size.

The \nlpwls{} reconstruction has improved resolution as compared to the
\begin{DIFmanualadd}\nlpwlsnc{}, \nlpwlsi{},\end{DIFmanualadd} and FDK reconstructions, with sharper trabecular bone boundaries.
Consequently, \nlpwls{} results in a more accurate trabecular segmentation\begin{DIFmanualadd}.
This is particularly evident when comparing to FDK and \nlpwlsi{}, where the segmentation images contain less detailed trabeculae\end{DIFmanualadd}.
This effect is well illustrated in the
Tb.Th.\ maps. The FDK and \begin{DIFmanualadd}\nlpwlsi{}\end{DIFmanualadd} maps show fewer, thicker
trabeculae, while the \nlpwls{} map is similar to the \uCT{}mv and \uCT\ maps with thinner and
more numerous trabeculae. \begin{DIFmanualadd}The \nlpwlsnc{} map is more similar to the \nlpwls{} map, but still contains thicker trabeculae.\end{DIFmanualadd} The mean Tb.Th.\ calculations (Table~\ref{tab:metrics}) confirm
this observation, with \nlpwls{} resulting in a Tb.Th.\ value closer to those of
\uCT{}mv and \uCT{} than do FDK\begin{DIFmanualadd}, \nlpwlsi{}, and \nlpwlsnc{}\end{DIFmanualadd}.
In contrast, \nlpwls{} shows no advantage with respect to Tb.Sp.\ and BV/TV.
BV/TV values are similar for \begin{DIFmanualadd}all methods, suggesting\end{DIFmanualadd} the loss of fine trabecular structures
and the increase in apparent trabecular thickness tend to cancel each other out in terms of BV/TV.
The same mechanism is a potential cause for the better accuracy of the FDK and \begin{DIFmanualadd}\nlpwlsi{}\end{DIFmanualadd} mean Tb.Sp. values:
the spacing lost to thicker trabeculae is recovered by the loss of fine trabeculae.
In contrast, \nlpwls{} does a better job in general at recovering small trabeculae, but
still reconstructs trabeculae as thicker than they should be, reducing the mean Tb.Sp.
Optimizing reconstructions based on one of these metrics instead of mJac may improve metric accuracy\begin{DIFtwo},
or show that \nlpwls{} is ill-suited to that metric\end{DIFtwo}.
\begin{table}
	\centering
	\caption{Trabecular bone metric results.}\label{tab:metrics}
	\begin{tabu} {X[l] X[r] X[r] X[r] X[r]}
		\toprule
		& mean Tb.Th. & mean Tb.Sp. & BV/TV \\
		& (mm) & (mm) & & \\
		\midrule
		FDK & \numr{3}{\fdkmean{}} & \numr{3}{\fdkSpmean} & \numr{3}{\fdkBV{}} \\
		\textcolor{diffcolor}{\nlpwlsi{}} & \textcolor{diffcolor}{\numr{3}{\nlpwlsspsImean{}}} & \textcolor{diffcolor}{\numr{3}{\nlpwlsspsISpmean}} & \textcolor{diffcolor}{\numr{3}{\nlpwlsspsIBV{}}} \\
		\textcolor{diffcolor}{\nlpwlsnc{}} & \textcolor{diffcolor}{\numr{3}{\nlpwlsspsBthreemean{}}} & \textcolor{diffcolor}{\numr{3}{\nlpwlsspsBthreeSpmean}} & \textcolor{diffcolor}{\numr{3}{\nlpwlsspsBthreeBV{}}} \\
		\nlpwls{} & \numr{3}{\nlpwlsspsmean{}} & \numr{3}{\nlpwlsspsSpmean} & \numr{3}{\nlpwlsspsBV{}} \\
		\uCT{}mv & \numr{3}{\uCTdsmean{}} & \numr{3}{\uCTdsSpmean} & \numr{3}{\uCTdsBV{}} \\
		\uCT{} & \numr{3}{\uCTmean{}} & --- & \numr{3}{\uCTBV{}} \\
		\bottomrule
	\end{tabu}
\end{table}

\section{Discussion}

We have presented a generalized reconstruction algorithm (\nlpwls{}) capable of utilizing a
variety of high fidelity CBCT system models, which may include focal-spot blur,
scintillator blur, and correlated noise. We evaluated this method in a
scintillator blur dominated scenario in simulation and on a prototype CBCT
test bench. These studies show that high fidelity modeling with \nlpwls\ can improve
resolution and produce more accurate reconstructions
as compared to \begin{DIFmanualadd}more traditional models\end{DIFmanualadd} and FDK approaches. The improved accuracy of the
trabecular bone segmentation and Tb.Th.\ measurement suggest that \nlpwls\ can
increase the accuracy of quantitative metrics used to study trabecular bone
health~\cite{mohan2013, Griffith2011, issever_assessment_2009}. Additionally, the improved \begin{DIFmanualadd}bias/noise\end{DIFmanualadd} trade-off \begin{DIFtwo}suggests\end{DIFtwo} that
	\nlpwls\ produces more accurate attenuation values \begin{DIFmanualadd}than \fdkdeblur{} and \nlpwlsi{}\end{DIFmanualadd}, which is critical for quantitative CT~\cite{cann_quantitative_1988} \begin{DIFtwo}(however, note that bias includes both attenuation value error and blurring)\end{DIFtwo}.

While this work utilizes a relatively simple mathematical formulation, we note that \nlpwls\ is
capable of incorporating a wide variety of complicated models. For example, one
can extend the model here to incorporate a shift-variant blur \begin{DIFmanualadd}and depth-dependence (focal-spot blur)\end{DIFmanualadd}~\cite{LaRiviere2007, tilley:16:msv} with proper
definition of $\mat{B}$ or $\mat{A}$. The model may also incorporate detector lag (a temporal
blur function) with a similar redefinition of $\mat{B}$ and blur due to gantry
motion with modifications to both $\mat{A}$ and $\mat{B}$. The only constraints are that $\mat{A}$, $\mat{B}$, and $\mat{W}$ 
are matrices and $\eta$ is positive. \begin{DIFmanualadd}Such modifications are the subject of ongoing and future work.\end{DIFmanualadd}

As algorithms enable increased resolution, proper choice of voxel size will be critical~\cite{sidky_constrained_2011}. 
If one were not attempting resolution recovery, the ideal voxel size would be about the size
of the demagnified system blur (\SI{0.33}{\milli\meter} for this system).
(The large system blur relative to pixel pitch results in most CBCT systems binning
projection data to increase effective pixel pitch.)
In this work voxel size was approximately equal to the demagnified pixel pitch (i.e., much smaller than the limit imposed by system blur).
Angular sampling also effects voxel size. CT data is almost always angularly undersampled.
To limit the effect of undersampling we acquire data in half angle increments (double the sampling of
traditional CBCT). %
In summary, we believe the choices of voxel size and angular sampling in this work are appropriate for the system blur studied,
and allow a fair comparison of the different MBIR system models.

While not a focus of this study, we note that incorporating blur into the model
decreases the convergence rate. In order to compare nearly converged solutions,
many iterations were used. (This is particularly important for regularization sweeps, as different regularization strengths may require different numbers of iterations.)
However, we believe that tuning the subset/acceleration
schedule can improve the convergence rate in practice. With the current (only partially optimized) implementation,
the bench data reconstructions took approximately \SIrange{10}{15}{\min} per iteration.
(Note the reconstruction volume was much larger than the ROI shown.)
When the ROI is small, as in this work, a multi-resolution reconstruction method may be employed to decrease iteration time~\cite{cao_multiresolution_2016}.

The main limitation of the objective function presented is the application of
the inverse covariance matrix, which may be computationally expensive if noise
correlations are modeled. In \begin{DIFmanualadd}the bench data study\end{DIFmanualadd}, we make assumptions to avoid computing this
inversion every iteration, but such assumptions will not always be valid \begin{DIFmanualadd}(as in the simulation study)\end{DIFmanualadd}. In
such cases, \begin{DIFmanualadd}one may need to make additional approximations to reduce computation time\end{DIFmanualadd}. Additionally, patient motion may be a resolution limiting factor
on high-resolution systems. However, if patient motion is
properly estimated, it may be incorporated into the system matrix to reduce this
image degradation without altering the presented algorithm~\cite{sisniega_motion_2017}.

The success of MBIR methods illustrates the importance of high-fidelity
modeling in CT reconstruction. Accurate modeling of CBCT systems, enabled by
the proposed method, improves image quality and permits high-resolution tasks
such as microcalcification detection and analysis of trabecular bone
morphology. In addition to improving the capabilities of current CBCT systems,
this method \begin{DIFmanualadd}has the potential to alter\end{DIFmanualadd} the trade-offs between hardware/geometry choices and image
quality, potentially effecting future CBCT system designs, including those that
aren't necessarily aiming for high resolution. For example, proper focal-spot
modeling may better leverage high-magnification or permit replacing
rotating-anode X-ray sources with more economical fixed-anode sources while
preserving resolution. Future studies will \begin{DIFmanualadd}characterize the improvements possible with the
proposed \nlpwls\ approach and their possible impact on CBCT system design, in addition to incorporating the different blur
models described above and considering systems with different balances of correlating
and non-correlating blur\end{DIFmanualadd}.

\section*{Acknowledgments}
This work was funded in part by NIH grant R21~EB014964, NIH grant
R01~EB018896, NIH grant F31~EB023783, and an academic-industry partnership with Varian Medical Systems
(Palo~Alto,~CA).
\begin{DIFmanualadd}Thanks to\end{DIFmanualadd} Yoshi Otake and Ali \"{U}neri for the GPU
software routines used in this work.
This work \begin{DIFmanualadd}used\end{DIFmanualadd} Maryland Advanced Research
Computing Center \begin{DIFmanualadd}resources\end{DIFmanualadd}.

\bibliography{ZoteroLibrarybibtex}

% Generated by IEEEtran.bst, version: 1.14 (2015/08/26)
\begin{thebibliography}{10}
\providecommand{\url}[1]{#1}
\csname url@samestyle\endcsname
\providecommand{\newblock}{\relax}
\providecommand{\bibinfo}[2]{#2}
\providecommand{\BIBentrySTDinterwordspacing}{\spaceskip=0pt\relax}
\providecommand{\BIBentryALTinterwordstretchfactor}{4}
\providecommand{\BIBentryALTinterwordspacing}{\spaceskip=\fontdimen2\font plus
\BIBentryALTinterwordstretchfactor\fontdimen3\font minus
  \fontdimen4\font\relax}
\providecommand{\BIBforeignlanguage}[2]{{%
\expandafter\ifx\csname l@#1\endcsname\relax
\typeout{** WARNING: IEEEtran.bst: No hyphenation pattern has been}%
\typeout{** loaded for the language `#1'. Using the pattern for}%
\typeout{** the default language instead.}%
\else
\language=\csname l@#1\endcsname
\fi
#2}}
\providecommand{\BIBdecl}{\relax}
\BIBdecl

\bibitem{Lai2007}
C.-J. Lai, C.~C. Shaw, L.~Chen, M.~C. Altunbas, X.~Liu, T.~Han, T.~Wang, W.~T.
  Yang, G.~J. Whitman, and S.-J. Tu, ``Visibility of microcalcification in cone
  beam breast {{CT}}: Effects of {{X}}-ray tube voltage and radiation dose.''
  \emph{Medical Physics}, vol.~34, no.~7, pp. 2995--3004, 2007.

\bibitem{Kwan2007}
A.~L.~C. Kwan, J.~M. Boone, K.~Yang, and S.-Y. Huang, ``Evaluation of the
  spatial resolution characteristics of a cone-beam breast {{CT}} scanner.''
  \emph{Medical Physics}, vol.~34, no.~1, pp. 275--281, 2007.

\bibitem{Carrino2014}
J.~A. Carrino, A.~Al~Muhit, W.~Zbijewski, G.~K. Thawait, J.~W. Stayman,
  N.~Packard, R.~Senn, D.~Yang, D.~H. Foos, J.~Yorkston, and J.~H. Siewerdsen,
  ``Dedicated cone-beam {{CT}} system for extremity imaging.''
  \emph{Radiology}, vol. 270, no.~3, pp. 816--24, 2014.

\bibitem{marinetto_quantification_2016}
E.~Marinetto, M.~Brehler, A.~Sisniega, Q.~Cao, J.~W. Stayman, J.~Yorkston,
  J.~H. Siewerdsen, and W.~Zbijewski, ``Quantification of bone
  microarchitecture in ultrahigh resolution extremities conebeam {{CT}} with a
  {{CMOS}} detector and compensation of patient motion,'' in \emph{Computer
  {{Assisted Radiology}} 30th {{International Congress}} and {{Exhibition}}},
  Heidelberg, Germany, Jun. 2016.

\bibitem{Gong2004}
X.~Gong, A.~A. Vedula, and S.~J. Glick, ``Microcalcification detection using
  cone-beam {{CT}} mammography with a flat-panel imager.'' \emph{Physics in
  Medicine and Biology}, vol.~49, no.~11, pp. 2183--2195, 2004.

\bibitem{Griffith2011}
J.~F. Griffith and H.~K. Genant, ``New imaging modalities in bone,''
  \emph{Current Rheumatology Reports}, vol.~13, no.~3, pp. 241--250, 2011.

\bibitem{baba_using_2004}
R.~Baba, K.~Ueda, and M.~Okabe, ``Using a flat-panel detector in high
  resolution cone beam {{CT}} for dental imaging,'' \emph{Dentomaxillofacial
  Radiology}, vol.~33, no.~5, pp. 285--290, Sep. 2004.

\bibitem{bamba_image_2013}
J.~Bamba, K.~Araki, A.~Endo, and T.~Okano, ``Image quality assessment of three
  cone beam {{CT}} machines using the {{SEDENTEXCT CT}} phantom,''
  \emph{Dentomaxillofacial Radiology}, vol.~42, no.~8, p. 20120445, Aug. 2013.

\bibitem{Thibault2007}
J.-B. Thibault, K.~D. Sauer, C.~A. Bouman, and J.~Hsieh, ``A three-dimensional
  statistical approach to improved image quality for multislice helical
  {{CT}},'' \emph{Medical Physics}, vol.~34, no.~11, p. 4526, 2007.

\bibitem{Feldkamp:84}
L.~A. Feldkamp, L.~C. Davis, and J.~W. Kress, ``Practical cone-beam
  algorithm,'' \emph{J. Opt. Soc. Am. A}, vol.~1, no.~6, pp. 612--619, Jun.
  1984.

\bibitem{Wang2014}
A.~S. Wang, J.~W. Stayman, Y.~Otake, G.~Kleinszig, S.~Vogt, G.~L. Gallia, A.~J.
  Khanna, and J.~H. Siewerdsen, ``Soft-tissue imaging with {{C}}-arm cone-beam
  {{CT}} using statistical reconstruction,'' \emph{Physics in Medicine and
  Biology}, vol.~59, no.~4, pp. 1005--1026, Feb. 2014.

\bibitem{Dang2015}
H.~Dang, J.~W. Stayman, A.~Sisniega, J.~Xu, W.~Zbijewski, X.~Wang, D.~H. Foos,
  N.~Aygun, V.~Koliatsos, and J.~H. Siewerdsen, ``Statistical
  {{Reconstruction}} for {{Cone}}-{{Beam CT}} with a
  {{Post}}-{{Artifact}}-{{Correction Noise Model}}: {{Application}} to
  {{High}}-{{Quality Head Imaging}},'' \emph{Physics in Medicine and Biology},
  vol.~60, no.~16, pp. 6153--6175, 2015.

\bibitem{Sun2015}
T.~Sun, N.~Sun, J.~Wang, and S.~Tan, ``Iterative {{CBCT}} reconstruction using
  {{Hessian}} penalty,'' \emph{Physics in Medicine and Biology}, vol.~60,
  no.~5, pp. 1965--1987, 2015.

\bibitem{Hofmann2013}
C.~Hofmann, M.~Knaup, and M.~Kachelriess, ``Do {{We Need}} to {{Model}} the
  {{Ray Profile}} in {{Iterative Clinical CT Image Reconstruction}}?'' in
  \emph{Radiological {{Society}} of {{North America Annual Meeting}}}, Chicago,
  2013, p. 403.

\bibitem{hofmann_effects_2014}
C.~Hofmann, M.~Knaup, and M.~Kachelrie{\ss}, ``\BIBforeignlanguage{en}{Effects
  of ray profile modeling on resolution recovery in clinical {{CT}}},''
  \emph{\BIBforeignlanguage{en}{Medical Physics}}, vol.~41, no.~2, pp.
  n/a--n/a, Feb. 2014.

\bibitem{Tsui1987}
B.~M.~W. Tsui, H.~B. Hu, D.~R. Gilland, and G.~T. Gullberg, ``Implementation of
  {{Simultaneous Attenuation}} and {{Detector Response Correction}} in
  {{Spect}}.'' \emph{IEEE Transactions on Nuclear Science}, vol.~35, no.~1, pp.
  778--783, 1987.

\bibitem{qi_fully_1998}
J.~Qi, R.~M. Leahy, C.~Hsu, T.~H. Farquhar, and S.~R. Cherry, ``Fully {{3D
  Bayesian}} image reconstruction for the {{ECAT EXACT HR}}+,'' \emph{IEEE
  Transactions on Nuclear Science}, vol.~45, no.~3, pp. 1096--1103, Jun. 1998.

\bibitem{qi_high-resolution_1998}
J.~Qi, R.~M. Leahy, S.~R. Cherry, A.~Chatziioannou, and T.~H. Farquhar,
  ``\BIBforeignlanguage{en}{High-resolution {{3D Bayesian}} image
  reconstruction using the {{microPET}} small-animal scanner},''
  \emph{\BIBforeignlanguage{en}{Physics in Medicine and Biology}}, vol.~43,
  no.~4, p. 1001, 1998.

\bibitem{formiconi_compensation_1989}
A.~R. Formiconi, A.~Pupi, and A.~Passeri,
  ``\BIBforeignlanguage{en}{Compensation of spatial system response in
  {{SPECT}} with conjugate gradient reconstruction technique},''
  \emph{\BIBforeignlanguage{en}{Physics in Medicine and Biology}}, vol.~34,
  no.~1, p.~69, 1989.

\bibitem{Chun2012}
S.~Y. Chun, J.~a~Fessler, and Y.~K. Dewaraja, ``Correction for
  collimator-detector response in {{SPECT}} using point spread function
  template,'' \emph{IEEE Transactions on Medical Imaging}, vol.~32, no.~2, pp.
  295--305, 2013.

\bibitem{alessio_modeling_2006}
A.~M. Alessio, P.~E. Kinahan, and T.~K. Lewellen, ``Modeling and incorporation
  of system response functions in 3-{{D}} whole body {{PET}},'' \emph{IEEE
  Transactions on Medical Imaging}, vol.~25, no.~7, pp. 828--837, Jul. 2006.

\bibitem{Alessio2003}
A.~Alessio, K.~Sauer, and C.~A. Bouman, ``{{MAP Reconstruction From Spatially
  Correlated PET Data}},'' \emph{IEEE Transactions on Nuclear Science},
  vol.~50, no.~5, pp. 1445--1451, 2003.

\bibitem{LaRiviere2006}
P.~J. La~Rivi{\`e}re, J.~Bian, and P.~A. Vargas, ``Penalized-likelihood
  sinogram restoration for computed tomography.'' \emph{IEEE Transactions on
  Medical Imaging}, vol.~25, no.~8, pp. 1022--36, Aug. 2006.

\bibitem{LaRiviere2007}
P.~J. La~Rivi{\`e}re and P.~Vargas, ``Correction for resolution nonuniformities
  caused by anode angulation in computed tomography,'' \emph{IEEE Transactions
  on Medical Imaging}, vol.~27, no.~9, pp. 1333--1341, 2008.

\bibitem{Zhang2014}
H.~Zhang, L.~Ouyang, J.~Ma, J.~Huang, W.~Chen, and J.~Wang, ``Noise correlation
  in {{CBCT}} projection data and its application for noise reduction in
  low-dose {{CBCT}}.'' \emph{Medical Physics}, vol.~41, no.~3, p. 031906, Mar.
  2014.

\bibitem{Yu2000}
D.~F. Yu, J.~A. Fessler, and E.~P. Ficaro, ``Maximum-likelihood transmission
  image reconstruction for overlapping transmission beams.'' \emph{IEEE
  Transactions on Medical Imaging}, vol.~19, no.~11, pp. 1094--105, Nov. 2000.

\bibitem{Feng2006}
B.~Feng, J.~A. Fessler, and M.~A. King, ``Incorporation of system resolution
  compensation ({{RC}}) in the ordered-subset transmission ({{OSTR}}) algorithm
  for transmission imaging in {{SPECT}}.'' \emph{IEEE Transactions on Medical
  Imaging}, vol.~25, no.~7, pp. 941--9, Jul. 2006.

\bibitem{zheng_Detector_2017}
J.~Zheng, J.~A. Fessler, and H.-P. Chan, ``Detector {{Blur}} and {{Correlated
  Noise Modeling}} for {{Digital Breast Tomosynthesis Reconstruction}},''
  \emph{IEEE Transactions on Medical Imaging}, 2017.

\bibitem{TilleyII2015a}
S.~Tilley~II, J.~H. Siewerdsen, and J.~W. Stayman,
  ``\BIBforeignlanguage{en}{Model-based iterative reconstruction for flat-panel
  cone-beam {{CT}} with focal spot blur, detector blur, and correlated
  noise},'' \emph{\BIBforeignlanguage{en}{Physics in Medicine and Biology}},
  vol.~61, no.~1, p. 296, 2016.

\bibitem{tilley_nonlinear_2016}
S.~Tilley~II, J.~H. Siewerdsen, W.~Zbijewski, and J.~W. Stayman, ``Nonlinear
  statistical reconstruction for flat-panel cone-beam {{CT}} with blur and
  correlated noise models,'' in \emph{{{SPIE}} 9783 {{Medical Imaging}} 2016:
  {{Physics}} of {{Medical Imaging}}}, vol. 9783, 2016, pp.
  97\,830R--97\,830R--6.

\bibitem{Erdogan1999}
H.~Erdo{\u g}an and J.~A. Fessler, ``Monotonic algorithms for transmission
  tomography.'' \emph{IEEE Transactions on Medical Imaging}, vol.~18, no.~9,
  pp. 801--14, Sep. 1999.

\bibitem{erdogan1999a}
H.~Erdo{\u g}an and J.~Fessler, ``Ordered subsets algorithms for transmission
  tomography,'' \emph{Physics in Medicine and Biology}, vol. 2835, 1999.

\bibitem{ding_quantification_2000}
M.~Ding and I.~Hvid, ``Quantification of age-related changes in the structure
  model type and trabecular thickness of human tibial cancellous bone,''
  \emph{Bone}, vol.~26, no.~3, pp. 291--295, Mar. 2000.

\bibitem{hildebrand_new_1997}
T.~Hildebrand and P.~R{\"u}egsegger, ``\BIBforeignlanguage{en}{A new method for
  the model-independent assessment of thickness in three-dimensional images},''
  \emph{\BIBforeignlanguage{en}{Journal of Microscopy}}, vol. 185, no.~1, pp.
  67--75, Jan. 1997.

\bibitem{Bouxsein2010}
M.~L. Bouxsein, S.~K. Boyd, B.~A. Christiansen, R.~E. Guldberg, K.~J. Jepsen,
  and R.~M{\"u}ller, ``Guidelines for assessment of bone microstructure in
  rodents using micro-computed tomography,'' \emph{Journal of Bone and Mineral
  Research}, vol.~25, no.~7, pp. 1468--1486, 2010.

\bibitem{TilleyII2016}
S.~Tilley~II, J.~H. Siewerdsen, and J.~W. Stayman, ``Nonlinear {{Statistical
  Reconstruction}} for {{Flat}}-{{Panel Cone}}-{{Beam CT}} with {{Blur}} and
  {{Correlated Noise Models}},'' in \emph{{{SPIE Medical Imaging}}}, San Diego,
  CA, 2016.

\bibitem{Nesterov2005}
Y.~Nesterov, ``Smooth minimization of non-smooth functions,''
  \emph{Mathematical Programming Journal, Series A}, vol. 103, pp. 127--152,
  2005.

\bibitem{Kim2015}
D.~Kim, S.~Ramani, and J.~A. Fessler, ``Combining {{Ordered Subsets}} and
  {{Momentum}} for {{Accelerated X}}-{{Ray CT Image Reconstruction}},''
  \emph{IEEE Transactions on Medical Imaging}, vol.~34, no.~1, pp. 167--178,
  2015.

\bibitem{nocedal_numerical_2006}
J.~Nocedal and S.~J. Wright, \emph{Numerical {{Optimization}}}, 2nd~ed., ser.
  Springer Series in Operation Research and Financial Engineering.\hskip 1em
  plus 0.5em minus 0.4em\relax New York, NY, USA: {Springer}, 2006.

\bibitem{hestenes1952methods}
M.~R. Hestenes and E.~Stiefel, ``Methods of conjugate gradients for solving
  linear systems,'' \emph{Journal of Research of the National Bureau of
  Standards}, vol.~49, no.~6, pp. 409--436, 1952.

\bibitem{Samei1998}
E.~Samei, M.~J. Flynn, and D.~A. Reimann, ``A method for measuring the
  presampled {{MTF}} of digital radiographic systems using an edge test
  device,'' \emph{Medical Physics}, vol.~25, no.~1, p. 102, 1998.

\bibitem{siewerdsen_signal_1998}
J.~H. Siewerdsen, L.~E. Antonuk, Y.~El-Mohri, J.~Yorkston, W.~Huang, and I.~A.
  Cunningham, ``Signal, noise power spectrum, and detective quantum efficiency
  of indirect-detection flat-panel imagers for diagnostic radiology,''
  \emph{Medical Physics}, vol.~25, no.~5, pp. 614--628, May 1998.

\bibitem{Long2010}
Y.~Long, J.~A. Fessler, and J.~M. Balter, ``{{3D}} forward and back-projection
  for {{X}}-ray {{CT}} using separable footprints.'' \emph{IEEE Transactions on
  Medical Imaging}, vol.~29, no.~11, pp. 1839--50, Nov. 2010.

\bibitem{jaccard_distribution_1912}
P.~Jaccard, ``\BIBforeignlanguage{en}{The {{Distribution}} of the {{Flora}} in
  the {{Alpine Zone}}.1},'' \emph{\BIBforeignlanguage{en}{New Phytologist}},
  vol.~11, no.~2, pp. 37--50, Feb. 1912.

\bibitem{huber_robust_statistics}
P.~J. Huber, \emph{Robust Statistics}.\hskip 1em plus 0.5em minus 0.4em\relax
  New York: {Wiley}, 1981.

\bibitem{klein_elastix:_2010}
S.~Klein, M.~Staring, K.~Murphy, M.~A. Viergever, and J.~P.~W. Pluim,
  ``Elastix: {{A Toolbox}} for {{Intensity}}-{{Based Medical Image
  Registration}},'' \emph{IEEE Transactions on Medical Imaging}, vol.~29,
  no.~1, pp. 196--205, Jan. 2010.

\bibitem{doube_bonej:_2010}
M.~Doube, M.~M. K{\l}osowski, I.~Arganda-Carreras, F.~P. Cordeli{\`e}res, R.~P.
  Dougherty, J.~S. Jackson, B.~Schmid, J.~R. Hutchinson, and S.~J. Shefelbine,
  ``{{BoneJ}}: {{Free}} and extensible bone image analysis in {{ImageJ}},''
  \emph{Bone}, vol.~47, no.~6, pp. 1076--1079, Dec. 2010.

\bibitem{schneider_nih_2012}
C.~A. Schneider, W.~S. Rasband, and K.~W. Eliceiri,
  ``\BIBforeignlanguage{en}{{{NIH Image}} to {{ImageJ}}: 25 years of image
  analysis},'' \emph{\BIBforeignlanguage{en}{Nature Methods}}, vol.~9, no.~7,
  pp. 671--675, Jul. 2012.

\bibitem{mohan2013}
G.~Mohan, E.~Perilli, I.~H. Parkinson, J.~M. Humphries, N.~L. Fazzalari, and
  J.~S. Kuliwaba, ``Pre-emptive, early, and delayed alendronate treatment in a
  rat model of knee osteoarthritis: {{Effect}} on subchondral trabecular bone
  microarchitecture and cartilage degradation of the tibia, bone/cartilage
  turnover, and joint discomfort,'' \emph{Osteoarthritis and Cartilage},
  vol.~21, no.~10, pp. 1595--1604, 2013.

\bibitem{issever_assessment_2009}
A.~S. Issever, T.~M. Link, M.~Kentenich, P.~Rogalla, A.~J. Burghardt, G.~J.
  Kazakia, S.~Majumdar, and G.~Diederichs, ``\BIBforeignlanguage{en}{Assessment
  of trabecular bone structure using {{MDCT}}: Comparison of 64- and 320-slice
  {{CT}} using {{HR}}-{{pQCT}} as the reference standard},''
  \emph{\BIBforeignlanguage{en}{European Radiology}}, vol.~20, no.~2, pp.
  458--468, Aug. 2009.

\bibitem{cann_quantitative_1988}
C.~E. Cann, ``Quantitative {{CT}} for determination of bone mineral density: A
  review.'' \emph{Radiology}, vol. 166, no.~2, pp. 509--522, Feb. 1988.

\bibitem{tilley:16:msv}
S.~Tilley~II, W.~Zbijewski, J.~H. Siewerdsen, and J.~W. Stayman, ``Modeling
  shift-variant {{X}}-ray focal spot blur for high-resolution flat-panel
  cone-beam {{CT}},'' in \emph{Proc. 4th {{Intl}}. {{Mtg}}. on Image Formation
  in {{X}}-Ray {{CT}}}, 2016.

\bibitem{sidky_constrained_2011}
E.~Y. Sidky, Y.~Duchin, X.~Pan, and C.~Ullberg, ``\BIBforeignlanguage{en}{A
  constrained, total-variation minimization algorithm for low-intensity x-ray
  {{CT}}},'' \emph{\BIBforeignlanguage{en}{Medical Physics}}, vol.~38, no.~S1,
  pp. S117--S125, Jul. 2011.

\bibitem{cao_multiresolution_2016}
Q.~Cao, W.~Zbijewski, A.~Sisniega, J.~Yorkston, J.~H. Siewerdsen, and J.~W.
  Stayman, ``\BIBforeignlanguage{en}{Multiresolution iterative reconstruction
  in high-resolution extremity cone-beam {{CT}}},''
  \emph{\BIBforeignlanguage{en}{Physics in Medicine and Biology}}, vol.~61,
  no.~20, p. 7263, 2016.

\bibitem{sisniega_motion_2017}
A.~Sisniega, J.~W. Stayman, J.~Yorkston, J.~H. Siewerdsen, and W.~Zbijewski,
  ``Motion compensation in extremity cone-beam {{CT}} using a penalized image
  sharpness criterion.'' \emph{Physics in Medicine and Biology (accepted)},
  2017.

\bibitem{DePierro1993}
A.~R. De~Pierro, ``On the relation between the {{ISRA}} and the {{EM}}
  algorithm for positron emission tomography,'' \emph{IEEE Transactions on
  Medical Imaging}, vol.~12, no.~2, pp. 328--333, 1993.

\bibitem{jacobson_expanded_2007}
M.~W. Jacobson and J.~A. Fessler, ``An {{Expanded Theoretical Treatment}} of
  {{Iteration}}-{{Dependent Majorize}}-{{Minimize Algorithms}},'' \emph{IEEE
  Transactions on Image Processing}, vol.~16, no.~10, pp. 2411--2422, Oct.
  2007.

\bibitem{prince_medical_2005}
J.~L. Prince and J.~Links, \emph{\BIBforeignlanguage{English}{Medical {{Imaging
  Signals}} and {{Systems}}}}, 1st~ed.\hskip 1em plus 0.5em minus 0.4em\relax
  Upper Saddle River, NJ: {Prentice Hall}, Apr. 2005.

\end{thebibliography}

\ifthenelse{\equal{\buildappendices}{true}}{
\clearpage

\appendices

\renewcommand{\theequation}{\Alph{section}.\arabic{equation}}

\begin{DIFmanualadd}\section{Update Step Derivation}\label{sec:opttransfer}\end{DIFmanualadd}

\setcounter{equation}{0}

Throughout this derivation, a lower case superscript in parenthesis denotes the current
iteration and a lower case subscript indicates an element of the vector \begin{DIFmanualadd}or\end{DIFmanualadd} matrix. For example, \begin{DIFmanualadd}$A_{ij}$ is the
element at the $i$\textsuperscript{th} row and $j$\textsuperscript{th} column of $\mat{A}$ and $l_i^{(n)}$ is the $i$\textsuperscript{th} element of $\vect{l}$ at iteration $n$\end{DIFmanualadd}. The subscripts $i$ and $j$ are used to index through the projection domain
and the image domain, respectively.

To simplify notation, let
\begin{equation}
        \vect{x} \triangleq \mx{}.
        \label{eq:x}
\end{equation}
To aid in deriving surrogates matched at the current iterate, $\theta$ may be expressed as
\begin{multline}
    \theta = \frac{1}{2} (\vect{x} - \vect{x}^{(n)})^T \mat{B}^T \mat{W} \mat{B} (\vect{x} - \vect{x}^{(n)}) + [\vect{x}^{(n)}]^T \mat{B}^T \mat{W} \mat{B} \vect{x} \\
    - \vect{y}^T \mat{W} \mat{B} \vect{x} - \frac{1}{2} [\vect{x}^{(n)}]^T \mat{B}^T \mat{W} \mat{B} \vect{x}^{(n)}. \label{eq:objective_x}
\end{multline}
A separable surrogate to $\theta$ may be found by replacing the Hessian ($\mat{B}^T \mat{W} \mat{B}$) with
$\mD\{\eta\}$ where
\begin{equation}
    \vect{\eta} \triangleq \mat{B}^T \mat{W} \mat{B} \vect{1}, \label{eq:eta}
\end{equation}
$\vect{1}$ is a vector of ones, and $\mD{}\{\cdot\}$ is
a diagonal matrix with its argument on the diagonal~\cite{DePierro1993, erdogan1999a, Kim2015}.  %
Thus, the quadratic surrogate may be expressed as
\begin{equation}
	Q_{X}^{(n)}(\begin{DIFmanualadd}{\vect{x}}\end{DIFmanualadd}) \triangleq \sum_i^{\begin{DIFmanualadd}n_y\end{DIFmanualadd}} \left ( \frac{1}{2} \begin{DIFmanualadd} x^2_i \eta_i + \rho_i^{(n)} x_i \end{DIFmanualadd} \right ) + \xi^{(n)} \label{eq:Qx}
\end{equation}
where
\begin{gather}
    \vect{\rho}^{(n)} \triangleq \mat{B}^T \mat{W} \mat{B} \vect{x}^{(n)} - \mD\{\vect{\eta}\} \vect{x}^{(n)} -
	\mat{B}^T \mat{W} \vect{y} \label{eq:rho}\\
	\xi^{(n)} \triangleq \frac{1}{2} \left [ [\vect{x}^{(n)}]^T \left ( \mD\{\vect{\eta}\} - \mat{B}^T \mat{W} \mat{B} \right ) \vect{x}^{(n)} \right]. \label{eq:xi}
\end{gather}
Note that $\xi^{(n)}$ is a constant, which can be ignored for the purposes of optimization.
Equation~\eqref{eq:Qx} is written using summation notation to highlight it's separability.
The first surrogate function, $Q$, is $Q_X$ expressed as a function of the line integrals $\vect{l}$:
\begin{multline}
	Q^{(n)}(\vect{l}) \triangleq Q^{(n)}_X(\mathrm{e}^{-\vect{l}}) = \sum_i^{\begin{DIFmanualadd}n_y\end{DIFmanualadd}} \begin{DIFmanualadd} Q^{(n)}_i = \end{DIFmanualadd} \\
	\begin{DIFmanualadd}\sum_i^{n_y} \frac{1}{2} \mathrm{e}^{-2 l_i} \eta_i + \mathrm{e}^{-l_i} \rho_i^{(n)} + \xi^{(n)} \end{DIFmanualadd} \label{eq:Q}
\end{multline}
where
\begin{equation}
    \vect{l} \triangleq \mat{A} \vect{\mu}.
\end{equation}
$Q_i^{(n)}$ is analogous to the marginal negative log-likelihood functions
in~\cite[Eq.~2]{Erdogan1999}. 

A quadratic surrogate to $Q_i^{(n)}$ is
\begin{multline}
    Q^{(n)}_{2,i}(\begin{DIFmanualadd}l\end{DIFmanualadd}_i) = Q^{(n)}_i(\begin{DIFmanualadd}l\end{DIFmanualadd}_i^{(n)}) +\\
    (\begin{DIFmanualadd}l\end{DIFmanualadd}_i - \begin{DIFmanualadd}l\end{DIFmanualadd}_i^{(n)})
    \frac{dQ^{(n)}(\begin{DIFmanualadd}l\end{DIFmanualadd}_i^{(n)})}{d\begin{DIFmanualadd}l\end{DIFmanualadd}_i} + \frac{1}{2} (\begin{DIFmanualadd}l\end{DIFmanualadd}_i - \begin{DIFmanualadd}l\end{DIFmanualadd}_i^{(n)})^2
	\begin{DIFmanualadd}c\end{DIFmanualadd}_i^{(n)}.
    \label{eq:Q2}
\end{multline}
Assuming that $\mat{B}$ and $\mat{W}$ are chosen such that $\vect{\eta}$ is positive, the
optimal curvatures (e.g., those producing the widest surrogate function and, 
thus, the largest step size) are~\cite[Eq. 28]{Erdogan1999}
\begin{equation}
	\begin{DIFmanualadd}c\end{DIFmanualadd}_i^{(n)} =
    \begin{cases}
        \left[ 2 \frac{Q^{(n)}_i(0) - Q_i^{(n)}(\begin{DIFmanualadd}l\end{DIFmanualadd}_i^{(n)}) + \begin{DIFmanualadd}l\end{DIFmanualadd}_i^{(n)}
        \frac{dQ^{(n)}_i}{d\begin{DIFmanualadd}l\end{DIFmanualadd}_i}({\begin{DIFmanualadd}l\end{DIFmanualadd}_i^{(n)}})}{(\begin{DIFmanualadd}l\end{DIFmanualadd}_i^{(n)})^2} \right]_+ & \begin{DIFmanualadd}l\end{DIFmanualadd}_i^{(n)} > 0
        \\
        {\left[  \frac{ d^2
        Q^{(n)}}{d(\begin{DIFmanualadd}l\end{DIFmanualadd}_i^{(n)})^2}(0) \right]_+} & \begin{DIFmanualadd}l\end{DIFmanualadd}_i^{(n)} = 0.
    \end{cases}
    \label{eq:curvature}
\end{equation}
Details of the optimal curvature derivation are given in
Appendix~\ref{sec:optc}.

Finally, a surrogate to $Q^{(n)}_2$ can be defined in a similar manner to
$Q^{(n)}_X$~\eqref{eq:Qx} and as shown in~\cite{erdogan1999a}:
\begin{equation}
	Q^{(n)}_2(\vect{l}) =
			\sum_i Q^{(n)}_{2,i}(\sum_j \begin{DIFmanualadd}A\end{DIFmanualadd}_{ij} \begin{DIFmanualadd}\mu\end{DIFmanualadd}_j)
	\le Q_3(\vect{\mu})
\end{equation}
where
\begin{gather}
		Q_3^{(n)}(\vect{\mu}) \triangleq \sum_{i,j} \frac{\begin{DIFmanualadd}A\end{DIFmanualadd}_{ij}}{\begin{DIFmanualadd}\gamma\end{DIFmanualadd}_{i}}
                    Q^{(n)}_{2,i}(\begin{DIFmanualadd}\gamma\end{DIFmanualadd}_i (\begin{DIFmanualadd}\mu\end{DIFmanualadd}_j - \begin{DIFmanualadd}\mu\end{DIFmanualadd}_j^{(n)})
						+ \begin{DIFmanualadd}\sum_j^{n_\mu}\end{DIFmanualadd}\begin{DIFmanualadd}A_{ij}\end{DIFmanualadd}\begin{DIFmanualadd}\mu\end{DIFmanualadd}^{(n)}_j) \\
	\vect{\gamma} \triangleq \mat{A} \vect{1}.\label{eq:gamma}
\end{gather}
This new function, $Q^{(n)}_3,$ is separable with respect to $\vect{\mu}$ 
and matches $\theta$ in function value and derivative
for $\vect{\mu} = \vect{\mu}^{(n)}$.

\begin{DIFmanualadd}
After obtaining a separable surrogate $\Phi$ for the penalty function
$\mR$ using~\cite{erdogan1999a} and~\cite{Erdogan1999}, the combined
surrogate of the full objective function~\eqref{eq:objective2} may be minimized. 
Using the following definitions for compactness:
\begin{alignat*}{2}
	\begin{DIFmanualadd}L\end{DIFmanualadd}_j^{(n)} \triangleq \frac{d Q^{(n)}_{3, j}}{d \begin{DIFmanualadd}\mu\end{DIFmanualadd}_j}(\vect{\mu}^{(n)}) &\quad
    \begin{DIFmanualadd}D\end{DIFmanualadd}_j^{(n)} \triangleq \frac{d^2 Q^{(n)}_{3, j}}{d \begin{DIFmanualadd}\mu\end{DIFmanualadd}_j^2}(\vect{\mu}^{(n)})\\
    \bigtriangledown \Phi_j^{(n)} \triangleq \frac{d \Phi^{(n)}_j}{d\begin{DIFmanualadd}\mu\end{DIFmanualadd}_j}(\vect{\mu}^{(n)})
    &\quad
    \bigtriangledown^2 \Phi_j^{(n)} \triangleq \frac{d^2 \Phi^{(n)}_j}{d\begin{DIFmanualadd}\mu\end{DIFmanualadd}_j^2}(\vect{\mu}^{(n)})
\end{alignat*}
the minimization is given
by
\begin{multline}
\argmin_{\begin{DIFmanualadd}\mu\end{DIFmanualadd}_j \ge 0} \, Q^{(n)}_{3,j}(\begin{DIFmanualadd}\mu\end{DIFmanualadd}_j) + \beta \Phi^{(n)}_j(\begin{DIFmanualadd}\mu\end{DIFmanualadd}_j) \\
= \left [ \begin{DIFmanualadd}\mu\end{DIFmanualadd}_j^{(n)} - \frac{\displaystyle \begin{DIFmanualadd}L\end{DIFmanualadd}_j^{(n)} + \beta \bigtriangledown \Phi_j^{(n)}
}{\displaystyle \begin{DIFmanualadd}D\end{DIFmanualadd}_j^{(n)}
+ \beta \bigtriangledown^2 \Phi_j^{(n)} } \right ]_+.
\end{multline}
Note that $\vect{\mu}$ is constrained to physically realistic values by the $[\cdot]_+$
operator, which is the maximum of its argument and zero.
All derivatives are evaluated at $\begin{DIFmanualadd}\mu\end{DIFmanualadd}_j = \begin{DIFmanualadd}\mu\end{DIFmanualadd}_j^{(n)}$, yielding
\begin{align}
\begin{DIFmanualadd}L\end{DIFmanualadd}_j^{(n)} &= \sum_i \begin{DIFmanualadd}A\end{DIFmanualadd}_{ij} \frac{d Q^{(n)}_{2, i}}{d
\begin{DIFmanualadd}l\end{DIFmanualadd}_i}(\begin{DIFmanualadd}l\end{DIFmanualadd}_i^{(n)})\\
	\begin{DIFmanualadd}\vect{L}\end{DIFmanualadd} &= \mat{A}^T (\begin{DIFmanualadd}\mD\{\vect{\eta}\}\end{DIFmanualadd} \mathrm{e}^{-2 \mat{A} \vect{\mu}^{(n)}} - \begin{DIFmanualadd}\mD\{\vect{\rho}^{(n)}\}\end{DIFmanualadd} \mathrm{e}^{-\mat{A} \vect{\mu}^{(n)}})\\
\begin{DIFmanualadd}D\end{DIFmanualadd}_j^{(n)} &= \sum_i \begin{DIFmanualadd}A\end{DIFmanualadd}_{ij} \begin{DIFmanualadd}\gamma\end{DIFmanualadd}_i
\frac{d^2 Q^{(n)}_{2,i}}{d \begin{DIFmanualadd}l\end{DIFmanualadd}_i^2}(\begin{DIFmanualadd}l\end{DIFmanualadd}_i^{(n)})
\\
	\begin{DIFmanualadd}\vect{D}\end{DIFmanualadd} &= \begin{DIFmanualadd}\mat{A}^T \begin{DIFmanualadd}\mD\{\vect{\gamma}\}\end{DIFmanualadd} \vect{c}^{(n)} \end{DIFmanualadd}
\end{align}
where $\vect{\eta}$, $\vect{\rho}$, $\vect{\gamma}$, and $\vect{c}$ are given in
Equations~\eqref{eq:eta},~\eqref{eq:rho},~\eqref{eq:gamma}, and~\eqref{eq:curvature}, respectively.

Because $Q_3^{(n)}$ is separable in $\vect{\mu}$, each $\begin{DIFmanualadd}\mu\end{DIFmanualadd}_j$ can be updated
simultaneously. This update step is the core iterative estimator shown in Algorithm~\ref{alg}.
It can be shown that the surrogate is jointly continuous in $\vect{\mu}$ and $\vect{\mu}^{(n)}$. Therefore, if the sequence
$\{\vect{\mu}^{(n)}\}$ generated using this update step has a limit, that limit is a stationary point of the objective function~\eqref{eq:objective}~\cite[Thm. 4.1]{jacobson_expanded_2007}.
\end{DIFmanualadd}

\section{Optimum Curvature Criteria}
\label{sec:optc}

\setcounter{equation}{0}

\renewcommand{\theenumi}{\Alph{enumi}}
\renewcommand{\theenumii}{\arabic{enumii}}
\renewcommand{\labelenumi}{}
\renewcommand{\labelenumii}{\theenumi\theenumii:}

Erdo\u{g}an and Fessler have shown that~\eqref{eq:curvature} is the optimum
curvature for a function $Q\begin{DIFmanualadd}(l)\end{DIFmanualadd}$ (see \begin{DIFmanualadd}Appendix\end{DIFmanualadd} of~\cite{Erdogan1999})
if:
\begin{enumerate}
\item \begin{enumerate}
      \item $\ddot{Q} > 0$ when $l \ge 0$.
      \item $\dddot{Q} < 0$ when $l \ge 0$.
      \end{enumerate}\label{it:A}
\item OR \begin{enumerate}
      \item $Q \in \mathbb{C}^2$.\label{it:C2}
      \item $\dot{Q}(l^*)$ is a local maximum of $\dot{Q}$.\label{it:max}
      \item $l^*$ is the only critical point of $\dot{Q}$.\label{it:crit}
      \item $\dddot{Q} < 0$ when $l < l^*$.\label{it:concave}
      \item $\ddot{Q} > 0$ when $l < l^*$.\label{it:mono}
      \end{enumerate}\label{it:B}
\end{enumerate}
We use dot notation to indicate derivatives with respect to $l$.

$Q$ is defined in~\eqref{eq:Q}. Without the subscripts and
superscripts,
\begin{equation}
Q(l) = \frac{1}{2} \eta \mathrm{e}^{-2l} + \rho \mathrm{e}^{-l} + k.
\end{equation}
The first, second, and third derivatives
are given by
\begin{gather}
\dot{Q} = -\eta \mathrm{e}^{-2l} - \rho \mathrm{e}^{-l}\label{eq:AdQ} \\
\ddot{Q} = 2\eta \mathrm{e}^{-2l} + \rho \mathrm{e}^{-l}\label{eq:AddQ}\\
\dddot{Q} = -4 \eta \mathrm{e}^{-2l} - \rho \mathrm{e}^{-l}.\label{eq:AdddQ}
\end{gather}

If either $\eta$ or $\rho$ is zero, or both $\eta$ and $\rho$ are positive,~\ref{it:A} is true.

We will show that if $\rho$ is negative and $\eta$ is positive,~\ref{it:B} is true.
By definition,~\ref{it:C2} is true.
$\ddot{Q}(l)$ is $0$ only when
\begin{equation}
l = l^* \triangleq -\log\left(\frac{-\rho}{2\eta}\right),
\end{equation}
satisfying~\ref{it:crit}.
\begin{equation}
\dddot{Q}(l^*) = \frac{-\rho^2}{2\eta} < 0,
\end{equation}
implying $\dot{Q}(l^*)$ is a maximum, and therefore~\ref{it:max}.
$\dddot{Q}$ only has one root at $l = -\log(-\rho / 4\eta)$, which is greater than $l^*$.
This combined with the fact that $\dddot{Q}(l^*) < 0$ implies~\ref{it:concave}.
Finally, because $l^*$ is the only root of $\ddot{Q}$ and $\dddot{Q}(l^*) <
0$,~\ref{it:mono} is satisfied.

\renewcommand{\theenumi}{\arabic{enumi}}
\renewcommand{\labelenumi}{\theenumi.}
\renewcommand{\labelenumii}{\theenumi\theenumii:}

\section{Test-Bench Gain Estimation}
\label{sec:gain}

\setcounter{equation}{0}

To accommodate the nonuniform illumination of the test-bench detector, photon flux was modeled with the following equation~\cite{prince_medical_2005}:
\begin{gather}
    \begin{DIFmanualadd}g\end{DIFmanualadd}_i = \begin{DIFmanualadd}p\end{DIFmanualadd}_i \begin{DIFmanualadd}f\end{DIFmanualadd}_i \label{eq:gainmodel} \\
    \begin{DIFmanualadd}f\end{DIFmanualadd}_i \sim \Poisson(I_0 \cos^3(\begin{DIFmanualadd}\theta\end{DIFmanualadd}_i))
\end{gather}
where $\begin{DIFmanualadd}g\end{DIFmanualadd}_i$ is a random variable representing an offset corrected gain (bare-beam) scan value for
pixel $i$, $I_0$ is a constant
value representing the photon flux at the piercing point, $\begin{DIFmanualadd}p\end{DIFmanualadd}_i$ is the detector
gain for pixel $i$ (detector units per photon), and $\begin{DIFmanualadd}\theta\end{DIFmanualadd}_i$ is
the angle between pixel $i$ and the piercing ray.
The parameters $I_0$ and $\begin{DIFmanualadd}p\end{DIFmanualadd}_i$ were estimated using the mean and noise properties of offset
corrected gain scans.
Measurements were corrected for detector gain (i.e., divided by $\begin{DIFmanualadd}p\end{DIFmanualadd}_i$) to obtain measurements in photon units. %
To account
for focal-spot intensity variations, a normalization factor $\begin{DIFmanualadd}n_v\end{DIFmanualadd}$ was calculated
for each frame ($\begin{DIFmanualadd}v\end{DIFmanualadd}$) using an unobstructed (bare-beam) region of the projection
data.
$\mat{G}$ was therefore a matrix which scaled the values of each pixel $i$ by $I_0
\cos^3(\begin{DIFmanualadd}\theta\end{DIFmanualadd}_i)$ and the values of each frame $\begin{DIFmanualadd}v\end{DIFmanualadd}$ by $\begin{DIFmanualadd}n_v\end{DIFmanualadd}$. 

\section{Forward Model Approximations}
\label{sec:convapprox}

\setcounter{equation}{0}

\newcommand{\omegamatrix}[1]{\begin{bmatrix} #1_1 \mat{I} & \hdots & #1_{n_k} \mat{I} \end{bmatrix}}
\newcommand{\omegamatrixb}[2]{\begin{bmatrix} #1_1 #2_1 & \hdots & #1_{n_k} #2_{n_k} \end{bmatrix}}
\newcommand{\omegamatrixc}[1]{\begin{bmatrix} #1_1 & \hdots & #1_{n_k} \end{bmatrix}}
\newcommand{\sourceletmatrix}[1]{\begin{bmatrix} #1_1 \\ \vdots \\ #1_{n_k} \end{bmatrix}}

\begin{table}
\caption{Summary of Notation}\label{tab:convapproxsymbols}
\begin{tabu}{c X[cm] c}
	\toprule
\textcolor{diffcolor}{Variable} & \textcolor{diffcolor}{\centering Description} & \textcolor{diffcolor}{Nominal Value} \\
 & & \textcolor{diffcolor}{or Size} \\
 \midrule
\textcolor{diffcolor}{$ d $} & \textcolor{diffcolor}{Subpixels per pixel} & \\
\textcolor{diffcolor}{$ \mat{S} $} & \textcolor{diffcolor}{Subpixel binning matrix} & \textcolor{diffcolor}{$n_y \times d n_y$}\\
\textcolor{diffcolor}{$ \tilde{\mat{B}} $} & \textcolor{diffcolor}{Scintillator blur matrix} & \textcolor{diffcolor}{$d n_y \times d n_y$} \\
\textcolor{diffcolor}{$ \tilde{\mat{G}}_k $} & \textcolor{diffcolor}{Gain term for sourcelet $k$} & \textcolor{diffcolor}{$d n_y \times d n_y$} \\
\textcolor{diffcolor}{$ n_k $} & \textcolor{diffcolor}{Number of sourcelets} & \\
\textcolor{diffcolor}{$ \tilde{\mat{A}}_k $} & \textcolor{diffcolor}{System matrix for sourcelet $k$.} & \\
\textcolor{diffcolor}{$ \mat{L}_k$} & \textcolor{diffcolor}{Shift matrix for sourcelet $k$.} & \textcolor{diffcolor}{$ n_y \times n_y$} \\
\bottomrule
\end{tabu}
\end{table}

\begin{DIFmanualadd}A full d\color{diffcolortwo}i\color{black}scr\color{diffcolortwo}e\color{black}tized (mono-energetic) forward model can be expressed as:
\begin{equation}
	\vect{y} = \mat{S} \tilde{\mat{B}} \omegamatrixc{\tilde{\mat{G}}} \exp(-\sourceletmatrix{\tilde{\mat{A}}} \vect{\mu})
\end{equation}
where new symbols are defined in Table~\ref{tab:convapproxsymbols}.
This model samples the focal spot into sourcelets and the pixels into subpixels.
The vector of attenuation values $\vect{\mu}$ is forward projected by each sourcelet (multiplied by each $\tilde{\mat{A}}_k$),
resulting in an individual subpixel line integral for each sourcelet.
The negative exponent of each line integral is taken, and the result is scaled by a sourcelet specific gain term $\tilde{\mat{G}}_k$ and summed over sourcelets.
A subpixel level blur is applied ($\tilde{\mat{B}}$), followed by conversion to pixel sized measurements by summing subpixels ($\mat{S}$).
While the presented algorithm is capable of incorporating this model directly, in this work we make a number of simplifying approximations.
Specifically, we don't model subpixels explicitly, and replace the multiple sourcelets model with a convolution operator.
We note that these approximations may not be appropriate for all imaging scenarios, especially those with large fields of view.

\subsection{Elimination of subpixels}
The separable footprints projector~\cite{Long2010} performs the integration over pixel area (i.e.~sum over subpixels)
inside the exponential.
We can therefore represent the separable footprints forward projection for sourcelet $k$ as
\begin{equation}
\mat{A}_k \triangleq \mat{S} \tilde{\mat{A}}_k.
\end{equation}
However, the blur matrix $\tilde{\mat{B}}$ requires an input of subpixel values.
An up-sampling matrix may be included to estimate subpixel values from the pixel values.
These approximations yield
\begin{equation}
\vect{y} \approx \mat{B}_d \omegamatrixc{\mat{G}} \exp(-\sourceletmatrix{\mat{A}} \vect{\mu})
		 \label{eq:detblurapprox}
\end{equation}
where
\begin{gather}
\mat{B}_d \triangleq \mat{S} \tilde{\mat{B}} \mat{S}^T d^{-1} \\
\mat{G}_k \triangleq \mat{S} \tilde{\mat{G}}_k \mat{S}^T d^{-1}
\end{gather}
and $d$ is the number of subpixels per pixel.
Note while we used $\mat{S}^T d^{-1}$ (the Moore-Penrose psuedoinverse of $\mat{S}$) to upsample the data, any upsampling matrix may be used for the derivation.
So long as $\tilde{\mat{B}}$ is circular and the upsampling/downsampling operations are shift-invariant, $\mat{B}$ represents
a convolution operator.

\subsection{Elimination of sourcelets}
At any given voxel, the collection of sourcelet system matrices may be approximated as a single
system matrix duplicated by the number of sourcelets, with each duplication left multiplied by $\mat{L}_k$,
which shifts the output in the detector plane. Starting from~\eqref{eq:detblurapprox}:
\begin{equation}
	\vect{y} \approx \mat{B}_d \omegamatrixc{\mat{G}} \exp(-\sourceletmatrix{\mat{L}} \mat{A} \vect{\mu}).
	\label{eq:srcdup}
\end{equation}
The duplication matrices are calculated based on the apparent sourcelet positions from a specific location within the field of view.
This approximation is therefore most appropriate near that location (in most cases isocenter),
and is generally accurate so long as the apparent focal spot size and shape are relatively constant throughout the field of view.

Equation~\ref{eq:srcdup} can be further approximated by moving the duplication/shift step outside of the exponential.
Additionally, we assume there is some $\mat{G}$ and set of $\omega_k$s such that
\begin{equation}
\mat{G} = \omega_k^{-1} \mat{G}_k
\end{equation}
for all $k$ (i.e., the $\mat{G}$ matrices are equivalent to within a scale factor).
\begin{equation}
	\vect{y} \approx \mat{B}_d \omegamatrix{\omega} \sourceletmatrix{\mat{G} \mat{L}} \exp(-\mat{A} \vect{\mu}).
	\label{eq:srcdup2}
\end{equation}
Because $\mat{G}$ may be locally approximated as a constant scaling, we can approximate~\eqref{eq:srcdup2} as
\begin{equation}
	\vect{y} \approx \mat{B}_d \omegamatrix{\omega} \sourceletmatrix{\mat{L}} \mat{G} \exp(-\mat{A} \vect{\mu}).
	\label{eq:srcdup3}
\end{equation}
Defining
\begin{equation}
\mat{B}_s \triangleq  \omegamatrix{\omega} \sourceletmatrix{\mat{L}}
\end{equation}
equation \eqref{eq:srcdup3} can be expressed as
\begin{equation}
\vect{y} \approx \mat{B}_d \mat{B}_s \mat{G} \exp (-\mat{A} \mu).
\end{equation}
Additionally, because each $\mat{L}_k$ is circulant (ignoring boundary conditions), $\mat{B}_s$ is the weighted sum of circulant matrices, and is therefore itself circulant.
Thus $\mat{B}_s$ represents a convolution matrix.
\end{DIFmanualadd}

}{}
\end{document}